\documentclass[manuscript]{acmart}
\usepackage{makecell}
\usepackage{multirow}
\usepackage{colortbl}
\usepackage{subcaption} 
\AtBeginDocument{%
  }

\setcopyright{acmlicensed}
\copyrightyear{2025}
\acmYear{2025}
\acmDOI{XXXXXXX.XXXXXXX}

\begin{document}

\title[``Tools or Friends?'' Personality-Aligned Writing Companions]{``Pragmatic Tools or Empowering Friends?'' Discovering and Co-Designing Personality-Aligned AI Writing Companions}


\author{Mengke Wu}
\email{mengkew2@illinois.edu}
\affiliation{
    \institution{University of Illinois Urbana-Champaign}
    \department{School of Information Sciences}
    \city{Champaign}
    \state{Illinois}
    \country{USA}
}

\author{Kexin Quan}
\email{kq4@illinois.edu}
\affiliation{
    \institution{University of Illinois Urbana-Champaign}
    \department{School of Information Sciences}
    \city{Champaign}
    \state{Illinois}
    \country{USA}
}

\author{Weizi Liu}
\email{weizi.liu@tcu.edu}
\affiliation{
    \institution{Texas Christian University}
    \department{Bob Schieffer College of Communication}
    \city{Fort Worth}
    \state{Texas}
    \country{USA}
}

\author{Mike Yao}
\email{mzyao@illinois.edu}
\affiliation{
    \institution{University of Illinois Urbana-Champaign}
    \department{Institute of Communications Research}
    \city{Champaign}
    \state{Illinois}
    \country{USA}
}

\author{Jessie Chin}
\email{chin5@illinois.edu}
\affiliation{
    \institution{University of Illinois Urbana-Champaign}
    \department{School of Information Sciences}
    \city{Champaign}
    \state{Illinois}
    \country{USA}
}

\renewcommand{\shortauthors}{Wu, et al.}

\begin{abstract}

The growing popularity of AI writing assistants presents exciting opportunities to craft tools that cater to diverse user needs. This study explores how personality shapes preferences for AI writing companions and how personalized designs can enhance human-AI teaming. In an exploratory co-design workshop, we worked with 24 writers with different profiles to surface ideas and map the design space for personality-aligned AI writing companions, focusing on functionality, interaction dynamics, and visual representations. Building on these insights, we developed two contrasting prototypes tailored to distinct writer profiles and engaged 8 participants with them as provocations to spark reflection and feedback. The results revealed strong connections between writer profiles and feature preferences, providing proof-of-concept for personality-driven divergence in AI writing support. This research highlights the critical role of team match in human-AI collaboration and underscores the importance of aligning AI systems with individual cognitive needs to improve user engagement and collaboration productivity.

\end{abstract}




\maketitle

\section{INTRODUCTION}

Writing is a deeply personal and cognitively demanding activity that varies in how individuals develop, communicate, and reflect on ideas, structure content, and manifest creativity \cite{levy2013science, deane2008cognitive}. Among the many factors that influence writing, personality plays a pivotal role—shaping not only writing style but also revision strategies and feedback interpretation \cite{jensen1984personality, Banaruee2017CorrectiveFA, Marefat2006STUDENTWP}. For example, writers with analytical or conscientious personalities often favor structured, detail-oriented compositions, while those with intuitive or empathetic personalities tend to prioritize flexibility, creativity, and emotional tone \cite {kufner2010tell, wolfradt2001individual, jensen1984personality}. The emergence of large language models (LLMs), such as Grammarly\footnote{https://www.grammarly.com/} and ChatGPT\footnote{https://chat.openai.com/}, has enabled a new generation of AI writing assistants. Research in Human-Computer Interaction (HCI) also highlights LLMs’ potential to offer a wide range of support for writers, from grammar and style suggestions to content generation and ideation (e.g., \cite{gero2022sparks, yuan2022wordcraft, clark2018creative, dang2022beyond}).

Alongside these technological developments, researchers increasingly explore the dynamic human-AI collaboration in writing contexts. Studies suggest that AI can augment human creativity as collaborative writing partners beyond mere tools \cite{ding2023mapping, yang2022ai, biermann2022tool}, yet, users hold diverse expectations of AI’s role as writing assistants—seeing them as collaborators, editors, or still utilitarian tools. Personality also emerges as a key factor driving these divergences, as users bring varied goals, writing habits, and interaction preferences to the table, especially in co-creative scenarios \cite{jakesch2023co, biermann2022tool, gero2019metaphoria}. Therefore, personalization has become an evolving emphasis in the development of these systems. However, personalization efforts in AI writing assistants often focus on internal algorithmic adaptation (e.g., \cite{tang2024step, gabriel2015inkwell}). Some studies have explored external design for personalization through controllable interfaces and supportive functions to adjust the writing process and outcomes (e.g., \cite{yeh2024ghostwriter, yuan2022wordcraft, shi2022effidit}), but these typically adopt a one-size-fits-all approach and lack deep user involvement in the design process. Limitations in accounting for individual differences in human-AI team matching can lead to mismatched expectations, reduced satisfaction, and disengagement, particularly in tasks that require cognitive and emotional congruence between user and system \cite{zhang2021ideal, huang2019human}.

Our research investigates the potential of personality-driven design for AI writing companions. Specifically, we seek to understand how personality influences collaborative writing processes and the associated design considerations. We ask: \textbf{How do personality profiles shape user preferences for AI writing companions, and what are the design priorities?} To structure this investigation, we formulated the following design questions:

\begin{itemize}
\item DQ1: How do writers with different personality profiles envision desirable functions, interaction dynamics, and visual representations for writing companions?
\item DQ2: Which features and representations receive consistent endorsement in each profile, and how do these outcomes refine priorities for personality-aware design?
\end{itemize}

\begin{figure}
    \centering
    \includegraphics[width=1\linewidth]{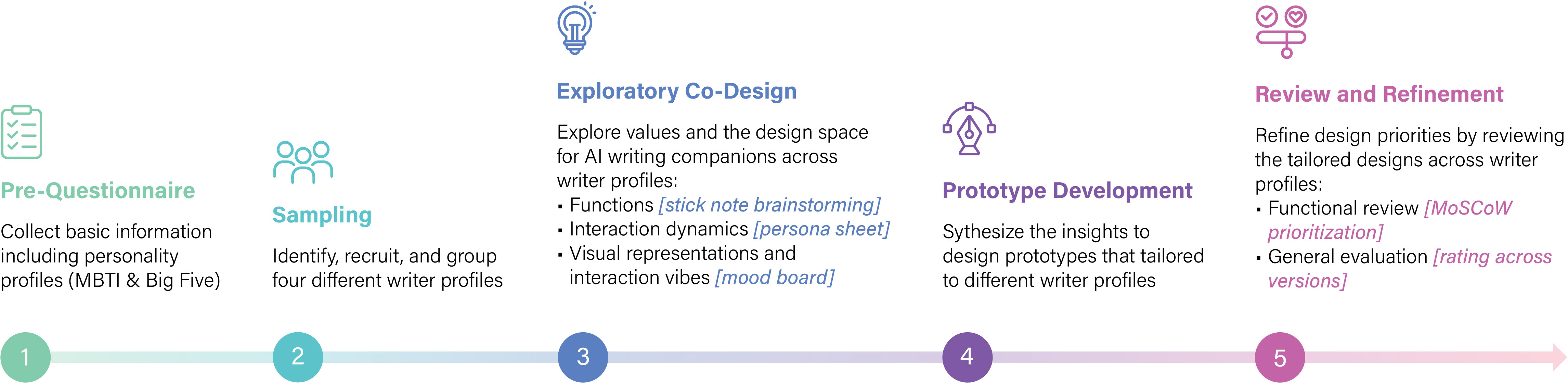}
    \caption{Overall Project Workflow: From Writer Profiling to Tailored Prototype Evaluation.}
    \label{fig:workflow}
\end{figure}

We conducted a multi-phase study to address these questions (see Figure \ref{fig:workflow} for an overview). First, we measured participants’ personality profiles using both the Big Five Inventory \cite{rammstedt2007measuring} and the Myers-Briggs Type Indicator (MBTI) \cite{myers2003mbti} before sampling. We used MBTI as a proxy to screen, recruit, and group participants into writer profiles, with different writing styles and preferences shaped by personality, and cross-validated these groupings with the Big Five. Drawing from prior research on writers' traits and AI usage \cite{gladis1993you, el2012adaptive, biermann2022tool, wang2024ai}, we used MBTI’s Sensing/Intuition (S/N) and
Thinking/Feeling (T/F) dimensions to explore how factors—such as attention to detail, emotional responsiveness, or thinking processes—might influence interactions with AI writing companions. In the first phase, we conducted \textbf{exploratory co-design workshops (DQ1)} with 24 participants to surface value-finding insights and elicit initial design ideas regarding desired functions, interaction dynamics, and visual representations of AI writing companions across writer profiles. Based on these insights, we developed two prototypes of AI writing companions, each embodied distinct personality-informed design philosophies. In the second phase, we conducted a \textbf{review and refinement workshop (DQ2)} and invited eight participants to engage directly with the prototypes. This stage was designed to assess participants’ feature preferences, confirm personality-aligned choices, and identify top-priority features to refine the design space.

Our work challenges the effectiveness of one-size-fits-all solutions in AI design, particularly in cognitively intensive human-AI collaboration like writing. By framing the study around these design questions, we not only map the design space but also apply personality-aware design principles to identify both shared and divergent expectations. This research contributes to:
\begin{itemize}
    \item Advancing discourse on human-AI collaboration by emphasizing the role of individual differences and the value of adaptive team matching.
    \item Providing empirical evidence on how personality influences user expectations, engagement, and satisfaction with AI, especially in co-creative writing contexts.
    \item Offering actionable insights for designing AI writing companions that move beyond surface-level personalization toward more effective, harmonious, and inclusive writing experiences.
\end{itemize}


\section{RELATED WORK}

\subsection{Individual Differences in Writing}\label{individual}
Writers differ in how they express themselves, shaped by a range of cognitive, psychological, sociocultural, and stylistic factors. Cognitive styles, for instance, affect working memory and planning strategies, influencing a writer’s ability to manage complex sentence structures and revise effectively \cite{flower1981cognitive, mccutchen2000knowledge}. Sociocultural and educational backgrounds also shape rhetorical structure and cohesion, as shown in contrastive rhetoric research and second-language writing studies \cite{kaplan1966cultural, silva1993toward}. Additionally, stylistic traits, such as function word choice, sentence length, and punctuation habits, tend to be stable enough for authorship attribution \cite{chung2011psychological, stamatatos2009survey}. Communication styles, studied in spoken discourse, similarly leave a mark on written tone and structure in expressiveness or preciseness \cite{ivanov2010behavioral}. These findings collectively reflect that writing is not merely a skill but a deeply personal act impacted by a series of individual characteristics.

Among these factors, personality plays an influential role. Defined as stable patterns of thinking, feeling, and behaving across time and contexts \cite{funder2012accurate}, personality not only affects social interactions but also leaves measurable traces in linguistic expression. Several frameworks exist for categorizing personality, with the Big Five and the Myers-Briggs Type Indicator (MBTI) being two widely applied models. The Big Five describes personality along five broad traits: Openness (intellectual curiosity and openness to try new things), Conscientiousness (self-discipline and to behave according to duty), Extraversion (pronounced engagement with the external world), Agreeableness (considerate, kind, and helpful with others), and Neuroticism (susceptibility to negative emotions like anger, anxiety, or depression) \cite{de2000big, rammstedt2007measuring}. These traits manifest in distinct writing behaviors and linguistic patterns. For instance, extraverted and agreeable people tend to use more social and positive-emotion words, while those high in openness display richer vocabulary and creativity \cite{pennebaker1999linguistic, kufner2010tell, wolfradt2001individual}. In educational contexts, personality also affects writing performance and strategy: introverts may produce more reflective and analytical writing, whereas extroverts often write expressively and engage more in narrative tasks \cite{he2019personality, tangkiengsirisin2020impact}. The MBTI, rooted in Jung’s theory of psychological types \cite{jung2016psychological}, offers another lens through which to understand writing behaviors. It classifies individuals into 16 profiles across four dichotomies: Extraversion/Introversion (E/I), Sensing/Intuition (S/N), Thinking/Feeling (T/F), and Judging/Perceiving (J/P) \cite{myers2003mbti}. These dimensions reflect cognitive styles influencing problem-solving, task structuring, and information management—factors directly relevant to writing. For example, writers differ in how they plan and organize ideas, with some favoring explicit, structured guidelines and step-by-step processes (S–J), while others work best when given the freedom to explore fluidly and abstractly (N–P); some prefer to explore thoughts through fast drafts (E) while others refine ideas internally before writing (I). Writing priorities also diverge: Thinkers (T) emphasize logical clarity and analytical precision, while Feelers (F) focus more on emotional resonance and audience connection \cite{jensen1984personality}. These tendencies reflect broader cognitive patterns: for example, \cite{el2012adaptive} observed that MBTI preferences, especially S/N and T/F, shape learners’ preferred modes of perceiving and structuring information, which likely extends to how they plan and compose written texts. Expanding on these insights, \cite{gladis1993you} identified four writer profiles: Correspondents (SF) for empathetic, audience-centered writing, Technical Writers (ST) for logical, precise content, Creative Writers (NF) for innovative, motivational pieces, and Analytical Writers (NT) for conceptual problem-solving.

Aforementioned discussions comprehensively suggest that personality not only influences writing style but also underpins the cognitive and motivational processes behind writing tasks. This insight has meaningful implications for AI-assisted writing systems: when the AI writing companion aligns its interaction style and feedback with a user’s personality-driven preferences, the writing experience could feel more intuitive and engaging. In this study, we examined how personality-informed writer profiles can guide the design of AI writing companions, offering practical frameworks for identifying user groups, analyzing writing behavior, and informing more personalized and effective human-AI writing collaboration.

\subsection{Human-AI Teaming in Writing}
The rise of AI has expanded opportunities for collaborative human–AI interactions, enabling humans to leverage AI tools for more efficient work. Research across education, healthcare, and business shows that AI collaboration models—where AI acts as both an assistant and co-cognitive partner—can improve outcomes \cite{park2019identifying, sowa2021cobots, wang2020human, kim2022learning}. These advances highlight the need for AI systems that complement human skills and adapt within teamwork settings, as also emphasized in team science research \cite{salas2008teams, andrews2023role}. In particular, human-AI teaming, a growing subfield, stresses close coordination between humans and AI, involving shared goals, mutually understood mental models, and aligned communication styles, to enhance decision-making, trust, and team performance \cite{zhang2021ideal, huang2019human, liang2019implicit}. Among collaborative applications, writing has become a prominent use case, with tools like ChatGPT widely adopted as writing assistants\footnote{https://sparktoro.com/blog/we-analyzed-millions-of-chatgpt-user-sessions-visits-are-down-29-since-may-programming-assistance-is-30-of-use/} \footnote{https://www.statista.com/statistics/1378998/chatgpt-use-tasks-us-by-type/}. This reliance motivates our investigation into analyzing and expanding the current design space in human-AI co-writing.

\subsubsection{Enhancing Writing through AI Collaboration}
Recent studies highlight the growing potential of AI as collaborative writing partners, supporting both practical and creative tasks—from fixed-scope assistance to complex, iterative co-creation requiring nuanced human–AI interaction \cite{ding2023mapping}. For example, \cite{yang2022ai} discovered a dynamic consideration of LLMs as active writers beyond a tool when users gain inspiration from AI in writing. \cite{clark2018creative} found that participants enjoyed co-writing with AI and were willing to reuse such systems, even though AI-assisted texts did not surpass human-authored ones from third-party evaluations. \cite{serban2017hierarchical} and \cite{gao2018neural} further explored AI’s role in storytelling and dialogue generation, suggesting that AI can actively participate in narrative construction. In educational contexts, AI has supported students’ writing through timely and adaptive feedback \cite{kangasharju2022lower}. Specifically, Large Language Models (LLMs) have proven effective in summarization \cite{dang2022beyond, goyal2022news}, content continuation and refinement \cite{lin2023and, yuan2022wordcraft}, and creative text generation \cite{mirowski2023co}, reflecting users’ openness to AI suggestions and their expectations for tools that blend direct edits with broader writing support. 

Complementing these system-oriented evaluations, user-centered studies reveal the complexities of trust and control in co-writing with AI. \cite{rezwana2022understanding} found that systems enabling two-way interaction improved user engagement and were viewed as more reliable and intelligent. Yet, \cite{liu2022will} demonstrated how users’ trust could diminish once they discovered that AI had authored the text. Other studies, such as \cite{jakesch2023co}, further revealed that writing with opinionated language models can shape users’ own attitudes, underscoring the persuasive power of AI-written text. These findings point to the need for transparency, explainability, and user agency in the design of AI writing tools to ensure not only usability but also ethical and trustworthy collaboration \cite{duin2023co, reza2025co}. Meeting these challenges requires more than technical improvements; it calls for active user empowerment throughout the design process. Designing AI writing tools with user input at the core ensures that systems reflect and respect users’ intentions—an alignment that can foster stronger trust and a greater sense of agency in human-AI partnerships.

\subsubsection{Personalization in AI Writing}
As AI writing tools become more integrated into everyday writing practices, the demand for personalization has grown. Users vary in their preferences when using AI tools, shaped by factors such as individual writing habits, goals, and deeper cognitive or emotional needs. For example, \cite{jakesch2023co} and \cite{xu2023chatgpt} indicate that user attitudes toward AI-generated content are influenced by the level of control they have, alignment of outputs with their personal views, and expectations for the writing task. Writers also want AI systems to respect their own values and writing preferences \cite{biermann2022tool}. Beyond practical concerns, writers’ engagement with AI collaborators is deeply shaped by a strong sense of ownership over their work \cite{gero2019metaphoria}. Ownership is not only tied to control over the creative process, but also emotional attachment and self-identity \cite{furby1978possession, dittmar1991meanings, belk1988possessions}. Writers, in particular, often develop deep connections toward their work in both creative and academic writing \cite{nicholes2017measuring}. Together with those influenced by personality traits (see Section \ref{individual}), these findings highlight the importance of designing AI writing tools that not only flexibly scaffold to individuals’ task preferences but also honor their needs for personal connection and creative autonomy.

Efforts to personalize AI writing tools often prioritize algorithmic adaptation, such as refining text generation by implicitly learning from user inputs (e.g., \cite{tang2024step, kim2019designing, salemi2023lamp}). For instance, \textit{InkWell} introduces stylistic variation generation by mimicking writers’ tone and personality to help refine their texts \cite{gabriel2015inkwell}. Beyond back-end adaptation, several studies have explored external system designs for AI writing tools, ranging from domain-specific applications—like language learning \cite{gayed2022exploring}, scientific writing and peer review \cite{kacena2024use, sun2024metawriter}, and creative storytelling \cite{yang2022ai}—to broader usability considerations that provide explicit means to improve usability, transparency, and user control \cite{yeh2024ghostwriter, yuan2022wordcraft, shi2022effidit}. For example, \textit{Wordcraft} supports story co-writing through a dialogue-based interface that allows users to interactively steer model outputs \cite{yuan2022wordcraft}. However, many efforts often adopt a one-size-fits-all approach, applying the same interfaces and mechanisms to all users. Moreover, most existing studies only conducted post-hoc usability assessments for the proposed co-writing tools but lack deep user engagement during the design process, thereby overlooking user-informed opportunities in the system design. Our work embraced a participatory approach to involve users early in the ideation stage to co-design personalized writing companions, capturing the diversity of their lived experiences, expectations, and preferences. This follows recent calls for more human-centered, user-involved design in AI system development \cite{fitzsimons2024overcoming, van2024exploring, delgado2023participatory}.

\subsection{Participatory Design Approach}
Participatory design (PD), or co-design, originated in Scandinavian labor movements in the 1970s and 1980s to involve workers in shaping workplace technologies \cite{schuler1993participatory}. In HCI, PD invites end-users and stakeholders as active collaborators throughout the design process to ensure the generated solutions align more closely with their real needs and goals \cite{kensing1998participatory, sanders2008co}. By leveraging diverse expertise, PD fosters empowerment and a sense of ownership over the final product \cite{sanders2002user}. This is particularly crucial in AI system development, where the opacity and developer control often limit user understanding and trust, as it enhances transparency and autonomy through active user participation \cite{zytko2022participatory}. Participatory design has also been successfully applied in fields like city planning, where local residents and officials collaboratively shape public spaces \cite{bratteteig2016unpacking}, and education, where teachers and students co-develop learning environments \cite{wright2010experience}. Overall, PD promotes more intuitive, equitable systems with deeper stakeholder involvement.


\section{EXPLORATORY CO-DESIGN WORKSHOP}

\subsection{Participants and Writer Profiles}
\begin{figure}[h]
    \centering
    \includegraphics[width=1\linewidth]{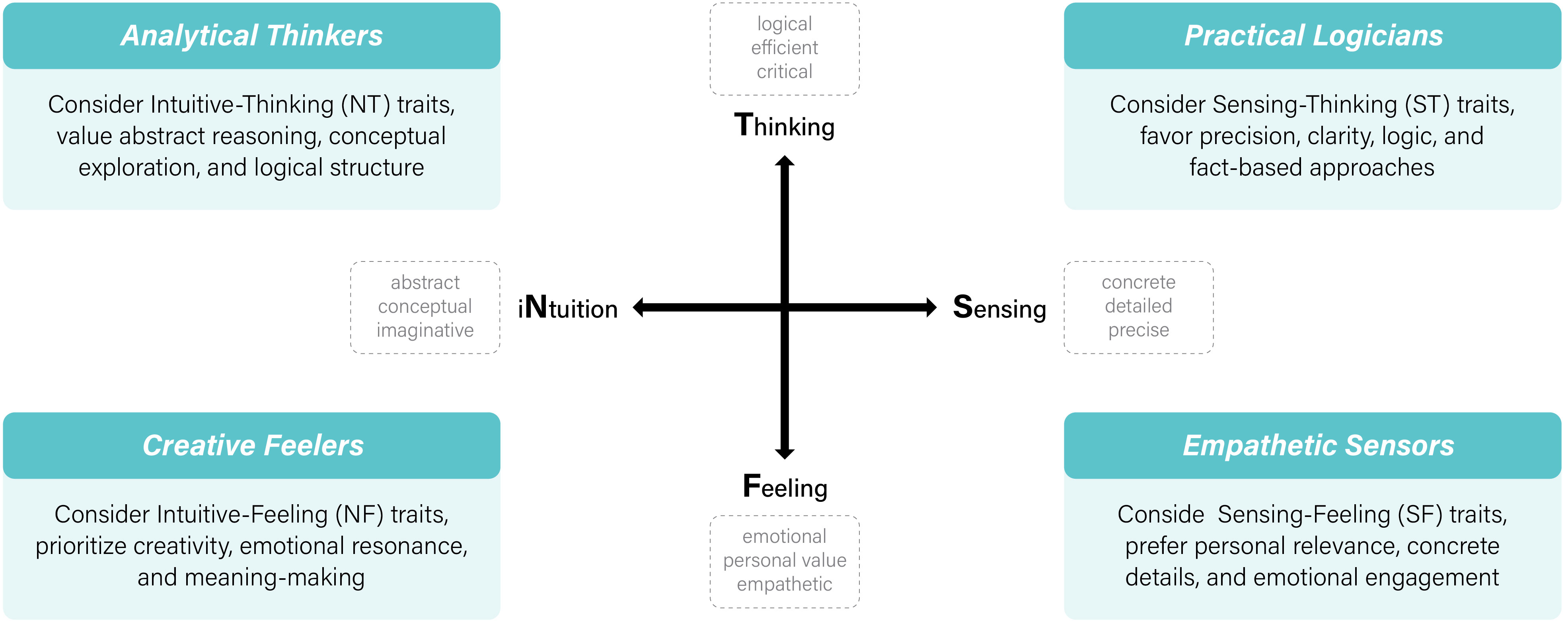}
    \caption{The Four Writer Profiles derived from Personality Traits.}
    \label{fig:profile}
\end{figure}

Participants were recruited from a U.S. public university via email lists and an online research pool, following IRB Approval. A pre-questionnaire collected demographic data, writing self-efficacy scores \cite{bruning2013examining}, prior design experience, and personality profiles using self-report MBTI and Big Five. We used MBTI as a proxy to identify and group distinct writer profiles based on their personality-driven cognitive and stylistic tendencies. Drawing from prior research on writer traits and AI usage \cite{gladis1993you, el2012adaptive, biermann2022tool, wang2024ai}, participants were recruited based on the Sensing/Intuition (S/N) and Thinking/Feeling (T/F) dimensions, which are particularly linked to writing styles, cognitive patterns, and interaction preference. The S/N dimension describes how individuals take in information: Sensing (S) focuses on concrete details and present realities, while Intuition (N) is drawn to abstract ideas and conceptual possibilities. The T/F dimension explains how people make decisions: Thinkers (T) rely on logic and objective criteria, whereas Feelers (F) are guided by empathy and personal values \cite{myers2003mbti}. Based on these traits, we operationalized four writer profiles (see Figure \ref{fig:profile} for details): 1) \textit{Analytical Thinkers} (considering NT traits), value abstract reasoning, conceptual exploration, and logical structures in writing; 2) \textit{Creative Feelers} (considering NF traits), prioritize creativity, emotional resonance, and meaning-making during creation; 3) \textit{Practical Logicians} (considering ST traits), favor precision, clarity, and fact-based approaches, and 4) \textit{Empathetic Sensors} (considering SF traits), prefer personal relevance, concrete details and emotional engagement. Given that Extraversion/Introversion (E/I) and Judging/Perceiving (J/P) primarily reflect social energy and organizational tendencies, which are less directly related to writing cognition, these dimensions were excluded in screening and grouping writer profiles.

To ensure a balanced representation across profiles, we also considered other factors from the pre-questionnaire when finalizing the sample. A total of 24 participants were recruited (six per writer profiles) and grouped into smaller workshops of three (with the same writer profiles) to ensure engagement of all participants. The participants' average age was 23.9 (\textit{SD} = 3.84), with 10 males, 13 females, and one genderqueer. Participants showed diverse writing self-efficacy (average = 72.9/100, \textit{SD} = 17.67). Eight participants had design experience, four had some, and nine had none (see Appendix \ref{appendix:profile} for more details). To further validate our MBTI-intiated writer groups, we examined their correlations and performances with participants’ Big Five traits \cite{rammstedt2007measuring}. The results showed alignment with MBTI’s S/N and T/F dimensions, providing converging evidence for the distinctiveness of the four writer profiles (see Section \ref{validation} for discussion).

\subsection{Workshop Materials and Procedure}
\begin{figure}[h]
    \centering
    \includegraphics[width=1\linewidth]{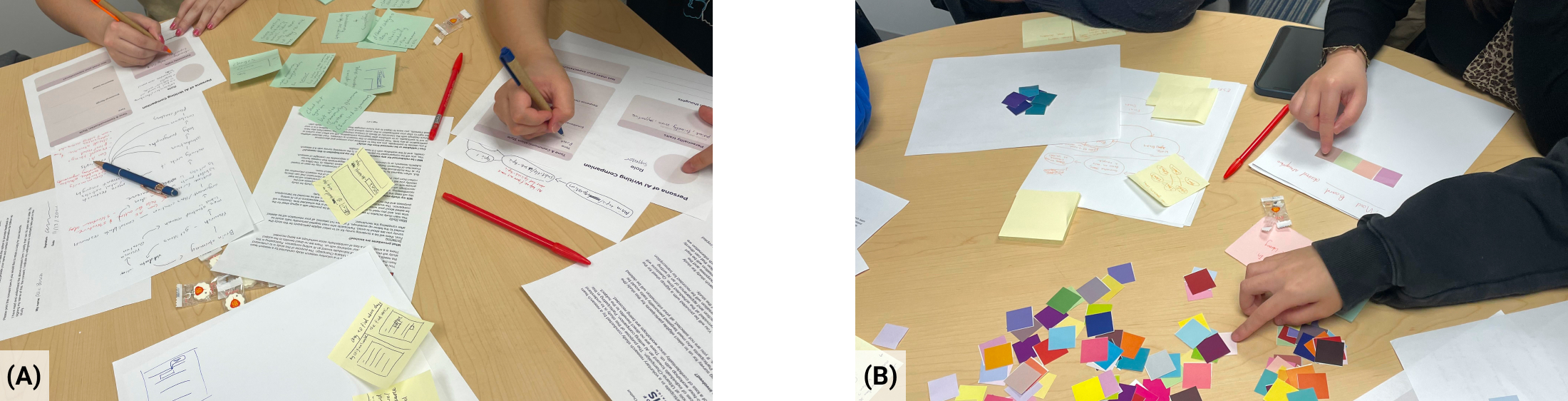}
    \caption{Example Workshop Activities: (A) Desired Function Brainstorming and Persona Construction, (B) Mood Board Creation.}
    \label{fig:photo}
\end{figure}

We employed participatory design to explore the design space for AI writing companions tailored to diverse personalities, capturing desired functions, traits, and visual representations. The 75-minute in-person workshops were structured to encourage both individual reflection and collaborative discussion. All workshops were recorded with participants' consent. The session began with an introduction to research objectives, participatory design methodology, and workshop structure. Participants then discussed their general writing experiences (e.g., what constitutes good writing, challenges in writing). Following this, they created flowcharts to illustrate their writing processes and show where and how Generative AI (GAI) usually assisted, followed by a group discussion. Next, participants entered a brainstorming session on features for ideal writing companions, documenting ideas on sticky notes that included suggestions to improve existing tools and new possibilities. These ideas were then shared in the group to identify common themes and develop new concepts. Participants then created persona sheets for their ideal companions, specifying preferences for their roles, personality traits, communication styles (tone, emotional range, and response detail), and how they should handle inaccuracies or misaligned suggestions (see Figure \ref{fig:photo}A). Finally, participants described their ideal companion’s appearance—such as human, animal, or object—and created mood boards \cite{lucero2012framing} to visually express their envisioned interaction vibes with the companion (see Figure \ref{fig:photo}B).

\subsection{Writer Grouping Validation}\label{validation}
This study leveraged established personality frameworks—specifically the MBTI and Big Five—to guide participant recruitment for writer profiles. Self-report MBTI served as the primary grouping proxy due to its intuitive structure, widespread adoption, and strong user familiarity, especially in digital spaces \cite{murphy2024artificial}. Its clear dichotomous categories make it practical for participant segmentation, while its popularity enhances engagement in user-centered research. MBTI’s widespread use in industry and professional contexts also lends external validity to its application in the design of AI writing companions \cite{mbti2025, wu2011application}.

We cross-validated our MBTI-initiated writer profiles against the Big Five. Previous research has documented strong correlations between the two models (e.g., \cite{harvey1995big, furnham1996big}): Extravertion/Introvertion (E/I) aligns with Big Five Extraversion, Sensing/Intuition (S/N) with Openness, Thinking/Feeling (T/F) with Agreeableness, and Judging/Perceiving (J/P) with Conscientiousness. These parallels suggest that MBTI captures meaningful aspects of personality. The correlational analysis supported the alignment between our writer profiles' traits and their corresponding Big Five dimensions (Table \ref{tab:cross}-1). Descriptive statistics also demonstrated distinct patterns across profiles (Table \ref{tab:cross}-2), reinforcing the validity of our classification. For example, participants in Feeling-oriented profiles (\textit{Creative Feelers} \& \textit{Empathetic Sensors}) scored higher on Agreeableness than those in Thinking-oriented profiles (\textit{Analytical Thinkers} \& \textit{Practical Logicians}). Similarly, profiles derived from Intuition (\textit{Creative Feelers} \& \textit{Analytical Thinkers}) exhibited higher Openness than those from Sensing (\textit{Empathetic Sensors} \& \textit{Practical Logicians}). Notably, Extraversion and Conscientiousness scores showed no significant variation across our writer profiles, supporting our balanced sampling for the E/I and J/P dimensions.

\begin{table}[h]
\centering
\begin{tabular}{
    >{\raggedright\arraybackslash}p{2.5cm}
    >{\centering\arraybackslash}p{2cm}
    >{\centering\arraybackslash}p{2cm}
    >{\centering\arraybackslash}p{2.5cm}
    >{\centering\arraybackslash}p{2cm}
    >{\centering\arraybackslash}p{2cm}
}
\toprule
 & \textbf{Extraversion} & \textbf{Agreeableness} & \textbf{Conscientiousness} & \textbf{Neuroticism} & \textbf{Openness} \\
\midrule
\multicolumn{6}{l}{\textbf{1 \textbar{} Correlation between the MBTI (row) and Big Five (column) dimensions for our samples}} \\[5pt]
E/I & 0.71\textsuperscript{***} & -0.20 & 0.00 & -0.36 & -0.43\textsuperscript{*} \\
S/N & -0.08 & 0.00 & -0.09 & 0.07 & 0.67\textsuperscript{***} \\
T/F & 0.18 & -0.53\textsuperscript{**} & 0.30 & -0.55\textsuperscript{**} & -0.22 \\
J/P & 0.28 & -0.05 & 0.65\textsuperscript{***} & -0.33 & -0.04 \\
\midrule
\multicolumn{6}{l}{\textbf{2 \textbar{} Mean (SD) of our writer profiles under the Big Five dimensions}} \\[5pt]
\textit{Analytical Thinkers} & 6.83 (1.47) & 6.00 (2.45) & 8.00 (1.79) & 4.00 (1.41) & 7.83 (0.98) \\
\textit{Creative Feelers} & 5.67 (2.34) & 8.50 (1.05) & 5.50 (2.07) & 7.50 (1.38) & 7.83 (1.33) \\
\textit{Practical Logicians} & 6.50 (1.38) & 6.67 (1.51) & 7.00 (1.26) & 5.17 (1.17) & 4.50 (2.07) \\
\textit{Empathetic Sensors} & 6.50 (1.64) & 7.83 (0.75) & 7.17 (2.14) & 5.83 (2.04) & 6.17 (0.75) \\
Sig Diff & \textit{p = 0.70} & \textit{p = 0.05} & \textit{p = 0.16} & \textit{p = 0.006} & \textit{p = 0.001} \\
\bottomrule
\end{tabular}
\begin{flushleft}
\footnotesize
\textit{Note}: *$p < .05$, **$p < .01$, ***$p < .001$.
\end{flushleft}
\caption{Cross-Validation with the Big Five for Our Writer Grouping.}
\label{tab:cross}
\end{table}



\section{EXPLORATORY DESIGN FINDINGS (DQ1)}\label{ideation}

The results of exploratory co-design workshop were transcribed and analyzed through thematic analysis \cite{braun2024thematic}. Two researchers independently coded the data, reconciled differences, and refined the codebook iteratively. The final code structure followed the workshop framework, including writing styles and preferences, desired features, interaction dynamics, and visual representations. Sub-themes emerged from these categories, such as \textit{“Writing/Response Style”} under desired features. Smaller topics, like \textit{"Creative Feelers: emotional appeal,"} were grouped as profile-specific content.
The findings are organized into three categories that align with the scope of DQ1: 1) functionalities, the features participants identified as desirable in writing companions; 2) interaction dynamics, the envisioned engagement and communication style with the companion; and 3) visual representations, the companion’s appearance, vibes, and interface design.

\subsection{Desired Functionalities for AI Writing Companions}

\begin{figure}
    \centering
    \includegraphics[width=1\linewidth]{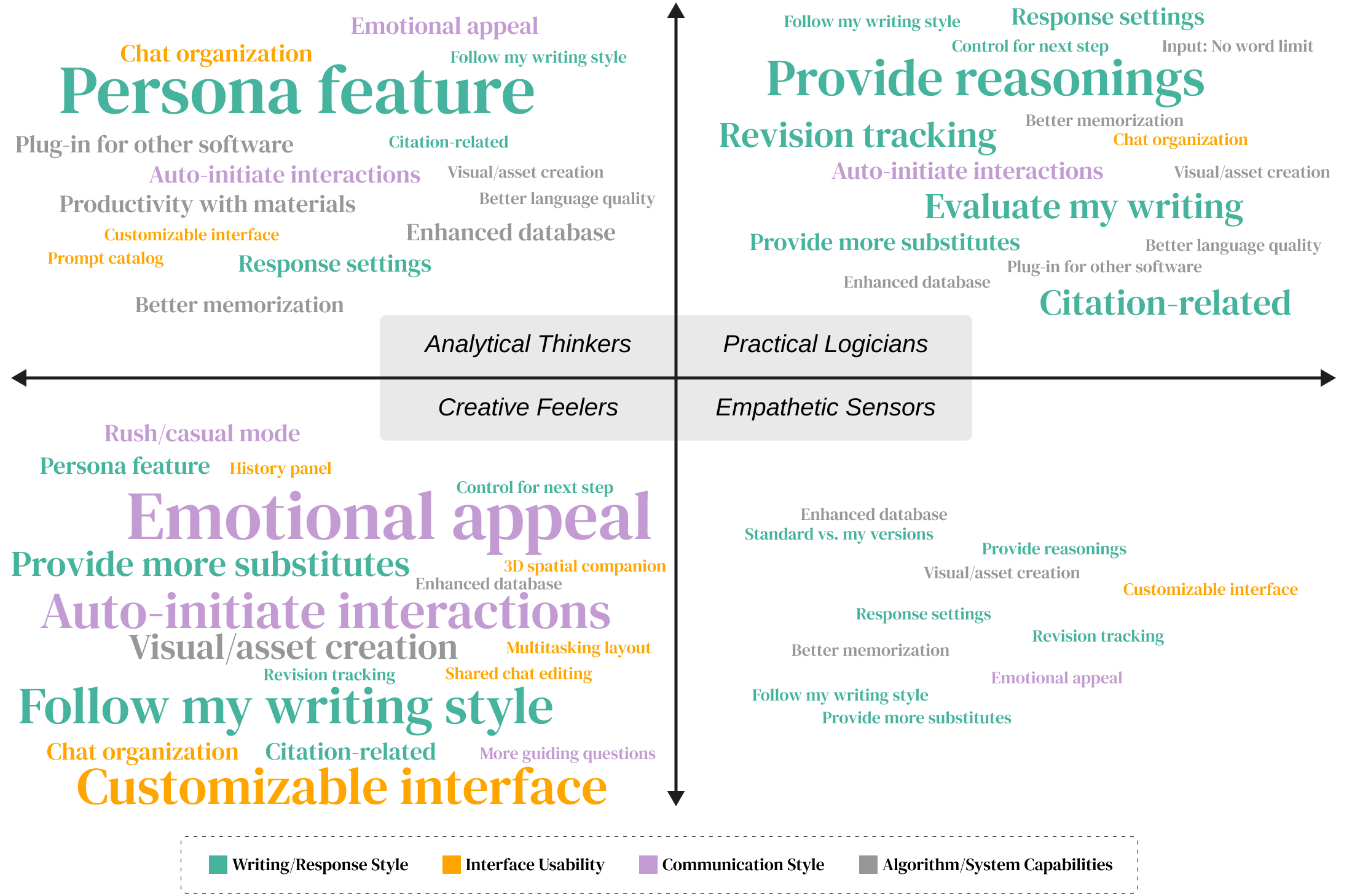}
    \caption{Frequency-Weighted Word Cloud of Key Features proposed by Different Writer Profiles.}
    \label{fig:wordcloud}
\end{figure}

This section outlines functional design space for AI writing companions, categorized into writing, interface, communication, and algorithm components. Figure \ref{fig:wordcloud} visualizes the key features proposed by writer profiles, with prominence indicating the frequency of participant mentions.

\subsubsection{Writing/Response Style} 
A central theme across participants (7 of 24) was the desire for the AI companion to \textbf{adapt to their individual writing styles} by remembering and referencing past interactions. Further suggestions included a slider to control how much the AI retains of their writing style (P9) and the ability to receive two output versions—one standard "good" AI version and one following the user’s personal style (P15). Building on this, the desire for a \textbf{persona feature} was particularly prominent among five Intuition participants (\textit{Analytical Thinkers} \& \textit{Creative Feelers}), reflecting their conceptual thinking processes. They thought it was inadequate in existing tools, like ChatGPT, where users input preferences in settings. P8 suggested a survey during account setup to allow the AI to learn more about the user. P7 even notably proposed a \textbf{switchable persona feature}, where users could set and choose from profiles based on specific needs, such as setting "student" for precision and clarity, "project manager" for structured, business-oriented communication, and "club organizer" for creative, engaging language. P3 also contributed to this thread by suggesting that, conversely, the companion could have personas to match different audience types (e.g., simpler language or slang for Gen Z).

\begin{figure}[b]
    \centering
    \includegraphics[width=1\linewidth]{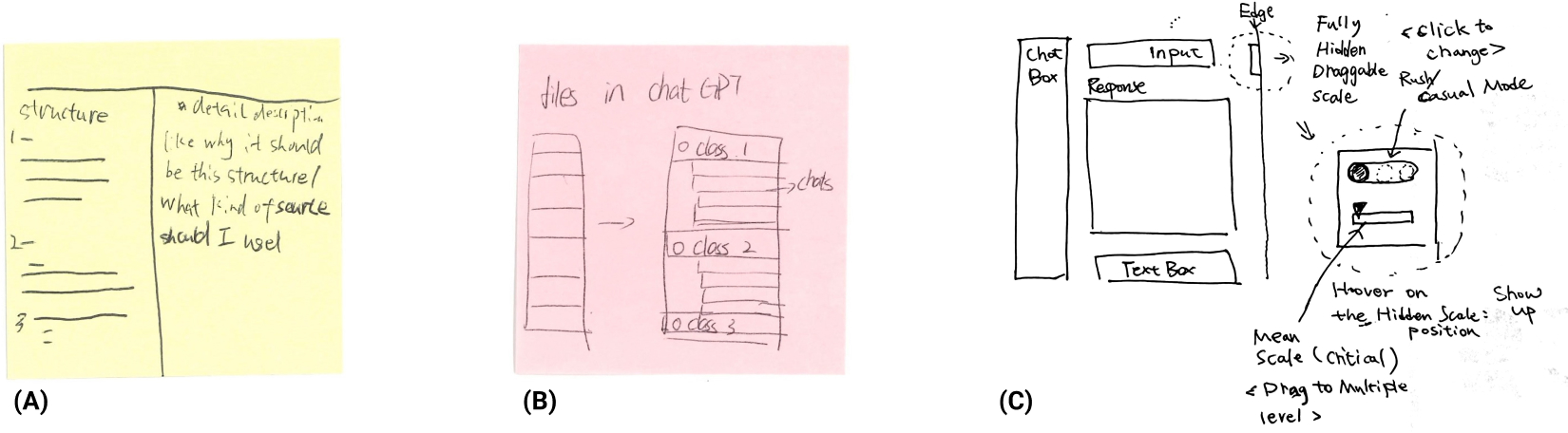}
    \caption{Example Design Sketches during the Workshop: (A) Providing Reasoning Side-by-Side, (B) Chat Organization, (C) Emotional Scale and Rush/Casual Mode.}
    \label{fig:draw}
\end{figure}

For Sensing and Thinking profiles (\textit{Practical Logicians} \& \textit{Empathetic Sensors} \& \textit{Analytical Thinkers}), \textbf{response customization} was a priority. Four expressed preferences for more control over the AI-generated response, such as tone (formal to casual), focus (e.g., clarity, grammar), context (e.g., academic, creative), and format (bullet points or paragraph). When asking for revisions, the detail-oriented Sensing profile (\textit{Practical Logicians} \& \textit{Empathetic Sensors}) emphasized better \textbf{revision tracking}. Two (P22, P23) praised Grammarly’s interface for clearly highlighting changes, with P23 proposing a toggle to bold/unbold changes. Additionally, P15 and P23 recommended a side-by-side comparison or a clickable button to review the original and revised versions. Transparency around revisions was also important to the Sensing profile, with four requesting \textbf{rationales for revisions} to explain AI suggestions. Layout ideas included a side-by-side panel (see Figure \ref{fig:draw}A), an explanation list at the end, hovering over changes to view reasons, and a toggle to show/hide reasons. Moreover, a feature resonated across all writer profiles—\textbf{the flexibility on receiving revision suggestions}. Seven participants desired the ability to request and review multiple alternatives for words or phrases in the given outputs (e.g., highlight and click to request).

Smaller but noteworthy features were also discussed. Participants raised the need for \textbf{citation-related} functionality. \textit{Practical Logicians} (P20, P24) advocated for auto-generated citation lists, with an option to toggle citations on or off. In contrast, \textit{Creative Feelers} (P4, P5) cared more about clarifying the credibility of sources, suggesting the inclusion of "verified labels" to separate trustworthy, traceable sources from AI-generated information. The fact-driven \textit{Practical Logicians} (P21, P22) also expressed interest in receiving \textbf{evaluations of their writing}, such as numerical grades, reading levels, and reading time.

\subsubsection{Interface Usability} This theme focuses on usability needs related to interface and visual elements, dominantly elaborated by Intuition and Feeling profiles (\textit{Analytical Thinkers} \& \textit{Creative Feelers} \& \textit{Empathetic Sensors}). Six of them emphasized the need for a \textbf{customizable interface} to match their cognitive and emotional preferences, such as altering colors and backgrounds for specific tasks or moods. Three \textit{Creative Feelers} (P4, P5, P6) further depicted task-based scenarios (e.g., plain for coding, vibrant for creative writing, and beach for travel itineraries) or time-based variations (e.g., “sky blue” in the morning, “orange” at lunchtime) to reflect shifts in energy and focus. P1 also re-imagined a multitasking layout to better organize the workspace, dividing the interface into blocks for the main task, thought trees, checklists, source lists, conversation with the companion, and a history panel, thus streamlining task organization and reducing cognitive load. P1 especially envisioned the \textbf{history panel} alike Photoshop that not only offers a navigational view of the entire workflow but also preserves records of all changes for users to experiment with companion interactions by jumping between steps.

Another key preference, particularly among the Intuition profile, was \textbf{chat organization} (see Figure \ref{fig:draw}B). P4 and P11 wanted to combine similar chats or organize them into custom collections for quicker access and easier management, such as by project, topic, or focus. Although only one \textit{Practical Logician} (P19) commented on chat organization, his suggestion stood out: an auto-generated \textbf{table of contents} with modifiable anchor points (similar to Google Docs) to improve within-chat navigation and topic organization for users who value precision and efficiency. Beyond chat, P7 also noted the need to organize prompts. She proposed a \textbf{“prompt catalog”}, as a library, to save or favorite frequently used prompts, streamlining repetitive typing tasks and prompt retrieval.

\subsubsection{Communication Style} Another popular topic was the communication style, such as the companion's conversational tone and dialogue structure when engaging with users. Insights from Intuition and Feeling profiles centered on \textbf{emotional appeal} (\textit{Creative Feelers} \& \textit{Empathetic Sensors}). Three suggested that the AI’s tone could be enriched with greater emotion and appreciated the support from anthropomorphic elements. However, \textit{Creative Feelers} also raised concerns about overly friendly communication. P5 explained, \textit{“I don't think the seemingly friendly tone really helps users improve their writing skills. Instead, it just violently changes [the writing] to a certain style.”} In response, three Feelers (P5, P6, P13) proposed a \textbf{"dragging emotional scale"} to adjust the companion’s critique level based on user needs (e.g., friendly to harsh) (see Figure \ref{fig:draw}C). P6 also suggested an option to disable emotional elements for users in a rush.

\textit{Creative Feelers} also provided unique input for dialogue style. For example, P5 wanted \textbf{more guiding questions} to encourage cognitive engagement. He noted that constantly and merely receiving answers \textit{“won’t become a knowledge of mine.”} Following this, P5 and P6 articulated a \textbf{“rush/casual mode”} (see Figure \ref{fig:draw}C). In “casual mode”, the companion would behave like a teaching assistant, prompting thoughtful discussions and reflection to let users find the answer step-by-step, while “rush mode” would prioritize efficiency to offer direct and concise answers without unnecessary conversations.

Across all writer profiles, there was broad support for the companion to take a proactive role and \textbf{auto-initiate interactions} rather than relying solely on users. However, their preferences varied. \textit{Creative Feelers} favored life and social connections, such as the companion greeting users with engaging messages when opening the app (P4), forwardly talking about context-aware suggestions (P2), and delivering timely reminders or alerts (P3). They also additionally proposed advanced \textbf{interaction paradigms} to deepen social and emotional attachments. P2 envisioned a spatial computing environment (e.g., AR), where the companion could \textit{“be a 3D buddy, standing there, understand my vision, and say ‘Hey, you make a mistake here!’”} to offer real-time, interactive feedback. In contrast, \textit{Practical Logicians} (P20) and \textit{Analytical Thinkers} (P9) focused on practical, task-oriented initiations from the companion. They valued a Word or Google Docs plug-in for contextually relevant ideas or feedback along with their thinking flows, rather than pasting content later after their mindset has shifted. P9 further recommended customizable thresholds for such auto-messages by frequency (e.g., every 10 mins) or severity (e.g., \textit{“only disturb me for significant problems”}), ensuring the support aligns with work needs and minimizes unnecessary interruptions. Beyond one-on-one communication, \textit{Creative Feelers} also suggested expanding interactions to multi-sided engagement, such as enabling users to \textbf{share chats for collaborative editing} (P6), akin to \textit{“group writing with AI,”} which reflected their conceptual and empathetic mindset.

\subsubsection{Algorithm and System Capabilities} Participants also called for backend technical features to enhance the companion's functionality, with input spanning across various personality types. A common request was for \textbf{better memorization} of interaction history to offer more personalized and accurate responses. They also sought \textbf{enhanced language quality}, with smoother phrasing, better paragraph connections, and avoidance of typical AI jargon that feels detached or robotic. A subtle inclination was identified for Intuition and Feeling profiles, who desired \textbf{visual/asset creation} tools like generating thought trees, idea webs, and diagrams to support their ideation and writing processes. Intuition profile was also keen on \textbf{better connection with external} for a more seamless writing workflow, such as integrating the companion as a plug-in for other applications or the ability to highlight cited parts in uploaded documents when generating answers.

\subsection{Desired Interaction Dynamics for AI Writing Companions}

Participants had distinct expectations for the \textbf{role} of writing companions, mirroring their personality traits. All \textit{Creative Feelers} viewed the companion as a motivator or supporter, focusing on emotional connection and encouragement to foster positive writing experiences. They imagined it as a "friend," with P2 even likening it to \textit{“a toy with a Mickey Mouse voice.”} Generally, \textit{Creative Feelers} preferred a warm, supportive companion with friendly, energetic, or empathetic tones for a creative and engaging environment. In stark contrast, all \textit{Practical Logicians} gravitated toward a more task-oriented, practical companion, seeing it as a "tool." They desired to retain the dominant role and positioned the companion as a passive assistant, follower, or information receiver. P21 even described his ideal companion as \textit{“an engine.”} For \textit{Practical Logicians}, emotional elements were negligible to professionalism, efficiency, and straightforwardness. Interestingly, \textit{Analytical Thinkers} and \textit{Empathetic Sensors}, who with more moderate personalities, correspondingly had more balanced and neutral expectations for the companion: aside from answers similar to those above, they delineated ideal companions as a friendly, patient, and reliable partner or collaborator, blending warmth with functionality.

The desired \textbf{conversation style} also varied. Intuition profile, particularly \textit{Creative Feelers}, uniquely valued guiding questions that stimulate deeper thinking, aligning with their preference for creative, reflective processes. Sensing profile (\textit{Practical Logicians} \& \textit{Empathetic Sensors}), conversely, preferred pragmatic, results-driven interactions with direct answers and clear reasoning. Regarding \textbf{companion's reactions to errors}, Feeling and Thinking profiles showed apparent differences again. Feelers (\textit{Creative Feelers} \& \textit{Empathetic Sensors}) wanted the companion to adopt an apologetic tone, whereas Thinkers (\textit{Practical Logicians} \& \textit{Analytical Thinkers}) firmly stated, \textit{“don’t apologize to me”} (P9, P24) but \textit{“explain what part is wrong”} (P8, P22, P23). Despite divergences, all profiles agreed that the companion should offer alternative suggestions that would help to resolve the issue.

\subsection{Desired Visual Representations for AI Writing Companions}

The \textbf{visual representation} of the writing companion also had striking differences, especially between Feelers and Thinkers. Eight Feelers (\textit{Creative Feelers} \& \textit{Empathetic Sensors}) were drawn to friendly and approachable designs and favored cute animals or anthropomorphic representations. Nevertheless, three Thinkers (\textit{Practical Logicians} \& \textit{Analytical Thinkers}) opted for robots, describing them as \textit{“closer to the machine nature”} (P8), or abstract shapes symbolizing intelligence and high-tech. Four Thinkers even explicitly preferred the existing text-based version without any decorative or dynamic elements, reinforcing their preference for simplicity and functionality. Notably, participants across all writer profiles shared reluctance toward human-like avatars, with comments like \textit{“very scary to me”} (P8) or \textit{“horrible”} (P15). When considering the \textbf{color preferences and overall vibes}, the divide between Feelers and Thinkers persisted. Feelers leaned toward vibrant, warm, or peaceful feelings that evoked emotional dynamics. Yet, Thinkers favored \textit{“simplistic, not too distracting”} (P19, P23) and prioritized muted, calm palettes or even just black-and-white, in line with their minimalist, no-frills approach.

\subsection{Synthesize Co-Design Findings}
\begin{figure}[h]
    \centering
    \includegraphics[width=1\linewidth]{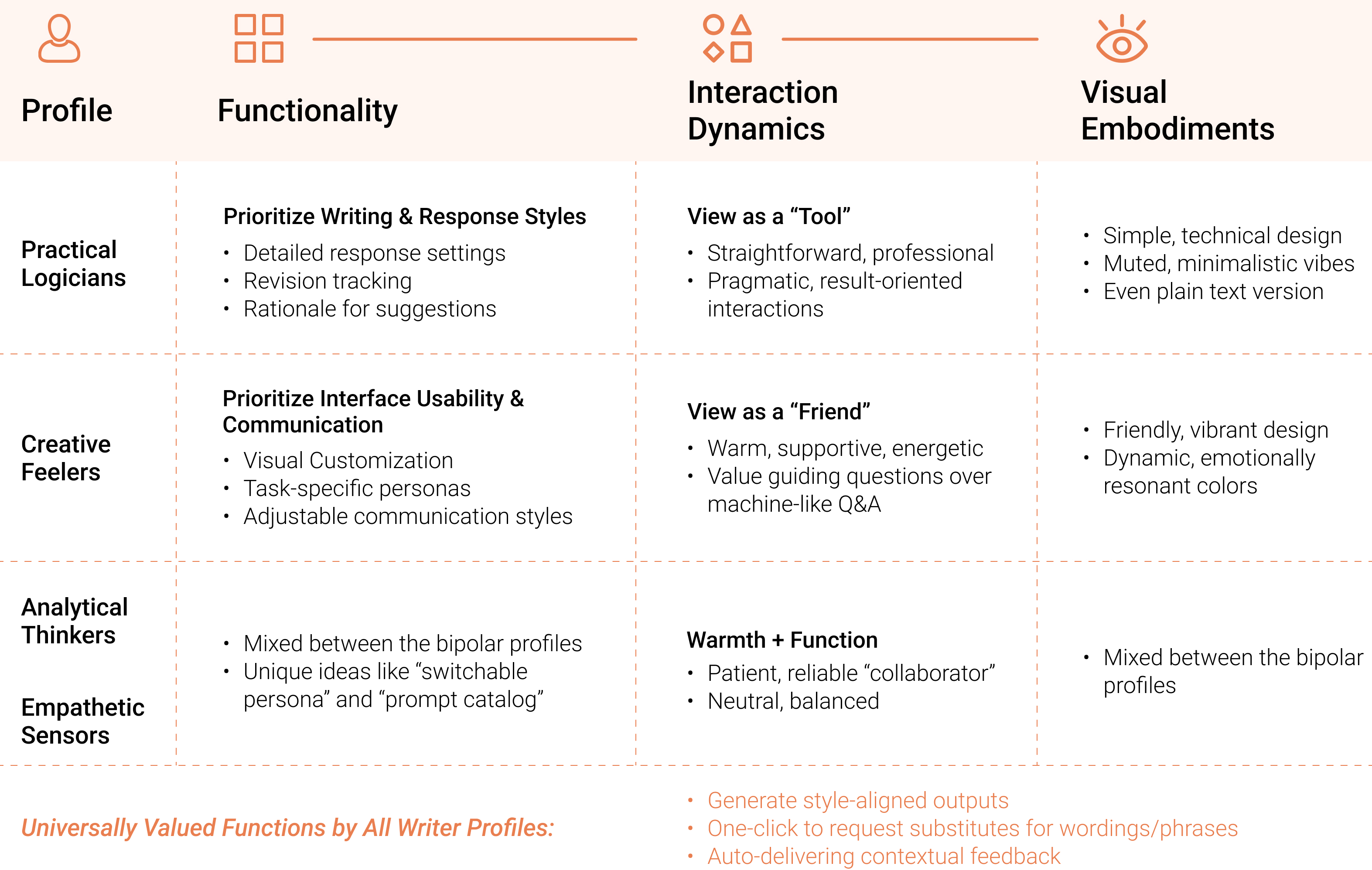}
    \caption{Comparative Summary of the Exploratory Design Preferences across Writer Profiles.}
    \label{fig:finding}
\end{figure}

Our co-design workshop uncovered design preferences for AI writing companions across different writer profiles. While participants agreed on a core set of useful features, clear divergences emerged that reflected distinct expectations and values. These differences underscore the role of personality in shaping design needs and point toward opportunities for customization to strengthen user engagement and human–AI teaming in writing (see Figure \ref{fig:finding}). 

\section{PROTOTYPE DESIGN AND IMPLEMENTATION}
Based on the values and ideas identified in Section \ref{ideation}, we conceived two design prototypes tailored to different cognitive styles and functional needs. These prototypes primarily focus on presenting core functions within typical usage flow, with complementary elements like conversational tone and visual design included to enrich the overall experience.

\subsection{Prototype Feature Selection}
First, we listed the proposed features by writer profiles to observe distribution trends and track how frequently each feature was mentioned. As discussed in Section \ref{ideation}, dominant divergence emerged between the bipolar \textit{Creative Feelers} and \textit{Practical Logicians}, while \textit{Analytical Thinkers} and \textit{Empathetic Sensors} tended to fall between, offering blended and overlapping ideas. To streamline design efficiency and avoid redundant or similar development, we consolidated the four profile probes into \textbf{a bipolar model}: “The Solution Master” (TSM) focuses on efficiency and structured reasoning that mainly informed by \textit{Practical Logicians}’ needs, and “The Empowering Pal” (TEP) promotes emotional connection and flexibility that mainly grounded in \textit{Creative Feelers}’ needs. Unique ideas from the moderate groups (\textit{Analytical Thinkers} \& \textit{Empathetic Sensors}) were carefully weighed for their alignments and contributions with each model and strategically assigned to the more fitting one. For features regarding algorithmic refinement, we omitted them in this stage due to technical constraints and our emphasis on UX features.

Based on the frequencies, we conducted team discussions to identify and prioritize main advocations that were commonly mentioned, carried strong user intent, and aligned with the overarching goal of either prototype. We selected an equal number of features in both prototypes for fair comparison. In total, 21 features were included: three shared functions across both, and nine specific to each. A complete function list with descriptions is provided in Table \ref{tab:function}.

Once the function sets were finalized, we revisited participants’ descriptions and usage scenarios to determine appropriate placements and interaction flows of each (e.g., on the landing page, when asking revisions). This review helped us understand when, how, and why users would engage with specific features during the writing process. Then, we moved from low- to high-fidelity design for the visual layout and interaction modes. Throughout the process, we iteratively tested the prototypes within the research team via repeated walkthroughs and interaction simulations to enhance the understandability and usability of the systems.

\begin{table}[h]
\centering
\resizebox{0.94\textwidth}{!}{
\begin{tabular}{l
>{\raggedright\arraybackslash}p{0.2\textwidth}
>{\raggedright\arraybackslash}p{0.5\textwidth}
>{\raggedright\arraybackslash}p{0.2\textwidth}}
\toprule
\textbf{ID} & \textbf{Name} & \textbf{Description} & \textbf{Location} \\
\midrule
\multicolumn{4}{l}{\textbf{Specific in "The Solution Master"}} \\[3pt]
TSM1 & Auto-feedback Settings & Set thresholds for auto-pop-up feedback (by frequency or severity, turn on/off). & system settings \\[14pt]
TSM2 & Response Settings & Adjust the tone, focus, context, wording, format, and length of the AI-generated writing. & prompt entry box \\[14pt]
TSM3 & Table of Contents & Auto-generate a table of contents within long chats for more efficient navigation and topic organization. & dialogue flow area \\[14pt]
TSM4 & Compare Before \& After & Revision tracking as bold or unbold changes and comparing the original and revised versions. & AI response \\[14pt]
TSM5 & Provide Reasoning & Provide rationales for revisions, with layout options as hover-to-see, side-by-side panel, and an end list. & AI response \\[14pt]
TSM6 & Prompt Catalog & Save or favorite frequently used prompts into a library for faster retrieval during conversation. & prompt entry box \\[14pt]
TSM7 & Slider: Keep \% of My Style & Adjust the retention percentage of the user’s writing style in generated outputs. & system settings \\[14pt]
TSM8 & Standard vs. My Style & Display two output versions: one standard "good" AI version vs. one reflecting the user’s writing style. & AI response \\[14pt]
TSM9 & Auto Citation List & Auto-generate citation lists at the end of the response. & AI response \\
\midrule
\multicolumn{4}{l}{\textbf{Specific in "The Empowering Pal"}} \\[3pt]
TEP1 & Auto-greeting Message & Greet users with interest-related messages or fun facts each day when they first open the companion. & main/landing page \\[14pt]
TEP2 & Visual Customization & Adjust the avatars, color themes, and background pictures of the companion. & system settings \\[14pt]
TEP3 & Switchable Persona & Select from multiple pre-set profiles based on their current needs (e.g., "student", "project manager"). & main/landing page, system settings \\[14pt]
TEP4 & Rush/Casual Mode & Adjust the dialogue style to be direct and efficient, or guiding, step-by-step, or reflective. & prompt entry box \\[14pt]
TEP5 & Emotional Scale & Adjust the companion’s critique level based on user goals or preferences (e.g., harsh to vibrant). & prompt entry box \\[14pt]
TEP6 & Chat Organization & Options to combine similar chats or organize them into custom collections. & chat history sidebar \\[14pt]
TEP7 & History Panel & A Photoshop-like history panel to outline the workflow and facilitate back-and-forth experimentation. & dialogue flow area \\[14pt]
TEP8 & "Verified label" for Reliable Sources & Special indicator for trustworthy sources to separate from AI-generated information. & AI response \\[14pt]
TEP9 & Collaborative Editing & Invite others to the same chat for joint writing (i.e., group editing with the AI companion). & dialogue flow area \\
\midrule
\multicolumn{4}{l}{\textbf{Mutual Functions in Both}} \\[3pt]
MF1 & Auto-pop-up Feedback & Provide context-aware suggestions during writing (e.g., plug-in in Word or Google Docs). & external application \\[13pt]
MF2 & Provide Substitutes & Highlight words or phrases to review substitutes. & AI response \\[3pt]
MF3 & Follow My Writing Style & Remember and refer to the user's prior inputs, interests, and writing samples to generate tailored outputs. & system settings \\
\bottomrule
\end{tabular}
}
\caption{Incorporated Functions in Each Prototype.}
\label{tab:function}
\end{table}

\subsection{Prototype Designs}

\subsubsection{"The Solution Master" (TSM)}
Designed for a \textbf{structured and efficient} writing experience, TSM primarily incorporated features from \textit{\textbf{Practical Logicians}}, complemented by structured ideas valued by \textit{Analytical Thinkers} and \textit{Empathetic Sensors}. It included three universally valued functions (MF1-3) and nine tailored ones (TSM1-9), strategically placed throughout the interface to support users who prioritize clarity, control, and precision. These tailored features were organized around several core design goals: enabling precise and well-aligned outputs (e.g., TSM2), enhancing transparency and fine revision (e.g., TSM4), and supporting efficient workflow and structured thinking (e.g., TSM3).

\begin{figure}[h]
    \centering
    \includegraphics[width=1\linewidth]{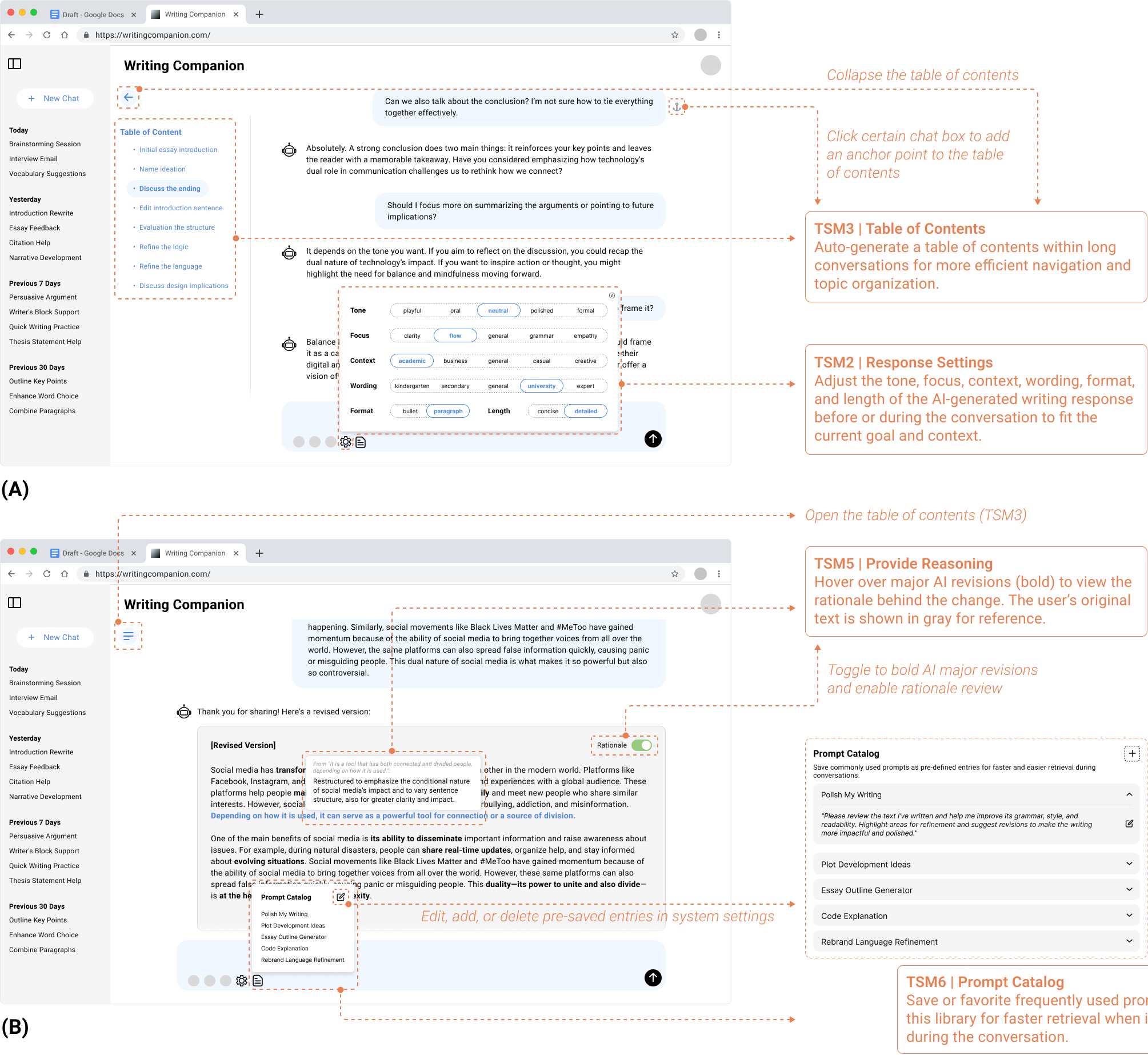}
    \caption{Representative Functions and Interfaces for \textbf{"The Solution Master" (TSM)}. (A) Normal Chat, Displaying “Response Settings” (TSM2) and “Table of Contents” (TSM3); (B) Ask for Revision, Displaying “Provide Reasoning” (TSM5) and “Prompt Catalog” (TSM6).}
    \label{fig:tsm}
\end{figure}

TSM's visual and interaction style also reinforced clarity and structure. It used a minimalist color palette, machine-like avatars, and a direct, calm tone of voice that prioritized answer delivery. Together, TSM supported high productivity and logically coherent writing with minimal friction. Table \ref{tab:function} lists all TSM functions, while Figure~\ref{fig:tsm} illustrates representative ones visually in the interface.

\subsubsection{"The Empowering Pal" (TEP)}

While TSM was built around precision and structure, TEP was designed for \textbf{warmth, creativity, and emotional connection} throughout the writing process. It drew primarily on preferences voiced by \textbf{\textit{Creative Feelers}}, enriched with supportive and expressive elements appreciated by \textit{Analytical Thinkers} and \textit{Empathetic Sensors}. Like TSM, TEP also included the three universally valued functions (MF1–3) and nine tailored ones (TEP1–9), but to foster emotional resonance, creative exploration, and compassionate AI collaboration. These tailored features were organized around: emotional and visual controls (e.g., TEP5), strengthening adaptability and trust (e.g., TEP3), and supporting organization and collaboration (e.g., TEP7). 

\begin{figure}[h]
    \centering
    \includegraphics[width=1\linewidth]{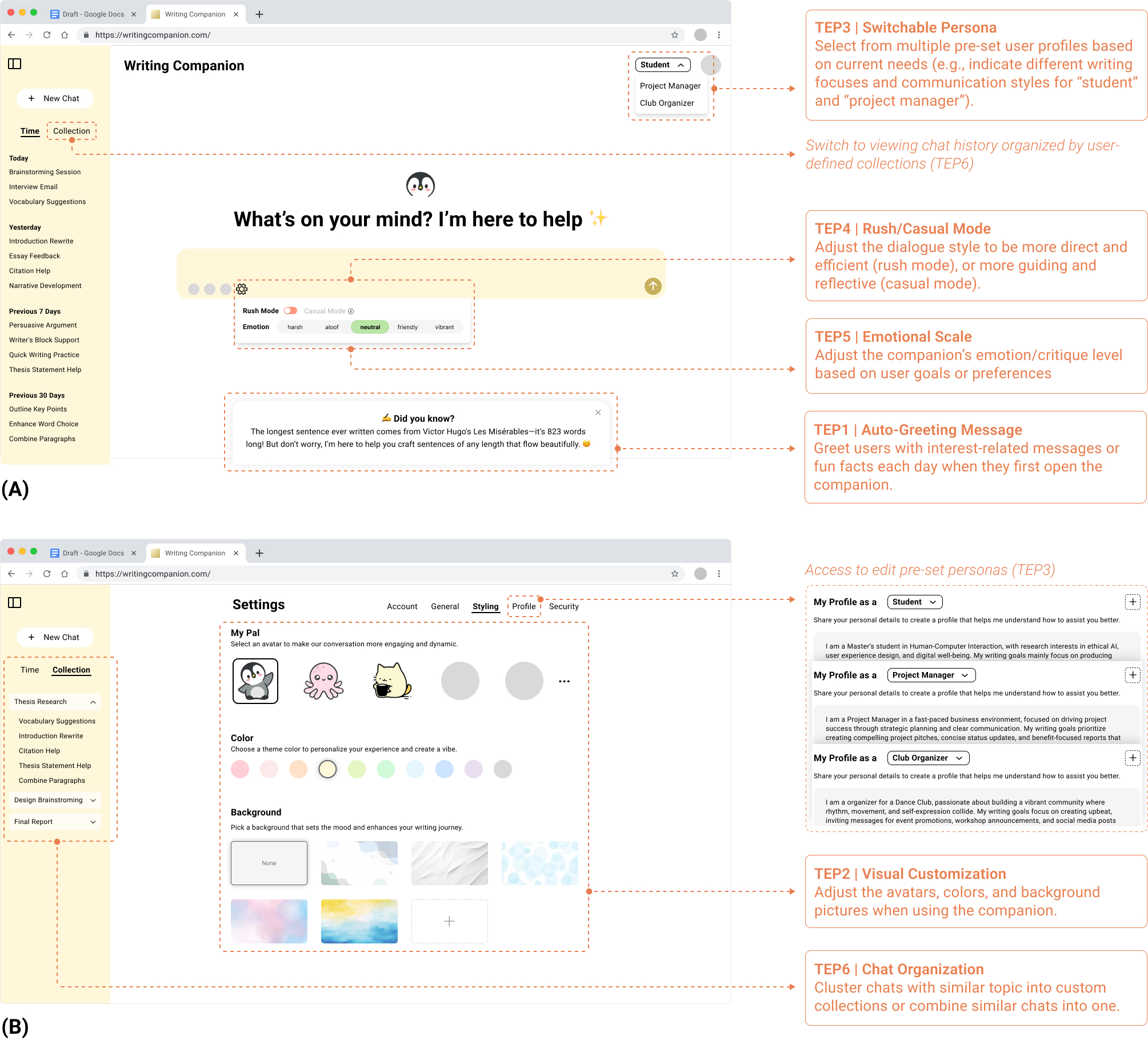}
    \caption{Representative Functions and Interfaces for \textbf{"The Empowering Pal" (TEP)}. (A) Landing Page, Displaying “Auto-Greeting Message” (TEP1), “Switchable Persona” (TEP3), “Rush/Casual Mode” (TEP4), and “Emotional Scale” (TEP5); (B) System Setting Page, Displaying “Visual Customization” (TEP2) and “Chat Organization” (TEP6).
}
    \label{fig:tep}
\end{figure}

Beyond functionality, TEP enhanced warmth and approachability through pastel colors, cute avatars (e.g., animated penguin), and vivid, encouraging conversational tones in the simulated dialogues. Rather than simply producing efficient outputs, TEP sought to uplift, co-create, and emotionally engage, turning writing into a thoughtful, expressive, and deeply human process. Table \ref{tab:function} lists all TEP functions, while Figure~\ref{fig:tep} illustrates representative ones visually in the interface.

\subsection{Implementation for Prototypes}

To balance user experience exploration with development feasibility, we employed a layered prototyping strategy built around a comprehensive visual foundation and supplemented by limited real-time interaction capabilities. The primary prototype was constructed using \textbf{Figma}, serving as the core platform to showcase the system's full functionality and design logic. To complement Figma’s static and pre-defined simulation flows, we also developed a lightweight \textbf{Executable Demo}—a web-based chatbot—specifically to illustrate the system’s content generation features in action.

This two-tiered approach allowed participants to evaluate both the system’s structural logic and experience flow (via Figma) and its text generation style and responsiveness (via the Executable Demo), ensuring a well-rounded and grounded understanding of the proposed AI writing companion.

\subsubsection{Figma Prototypes as Core} The Figma prototypes served as the primary representation of each system, visually displaying the full range of features outlined in Table 1 (see Figure \ref{fig:tsm} and \ref{fig:tep}). These high-fidelity, interactive mockups supported clicking on all essential chips and other interactive gestures (e.g., auto-pop-up messages, hover to view rationale, scroll through simulated conversations), effectively conveying the companion’s visual structure and user flow. This approach ensured a coherent and immersive experience, allowing participants to comprehend the system’s intended design without requiring underlying code execution.

\subsubsection{Executable Demo as Auxiliary} Given Figma's inherent limitation as a static prototyping tool, it cannot simulate real-time response generation. We thereby developed an Executable Demo as a targeted auxiliary. This simplified, web-based chatbot was built using React.js (TypeScript) and integrated with OpenAI’s API (default gpt-4o-mini) to demonstrate how the content-related features behave during live interaction. Unlike the Figma prototypes, the Executable Demo did not aim to mirror the full layout or navigation flow, but rather to demonstrate the output of specific features that rely on dynamic text generation. During user testing, this distinction was clearly communicated: the Figma prototypes were the primary reference for interface design and system flow, while the Executable Demo functioned solely as a supporting tool to illustrate content-based functions.

\begin{figure}[h]
    \centering
    \includegraphics[width=1\linewidth]{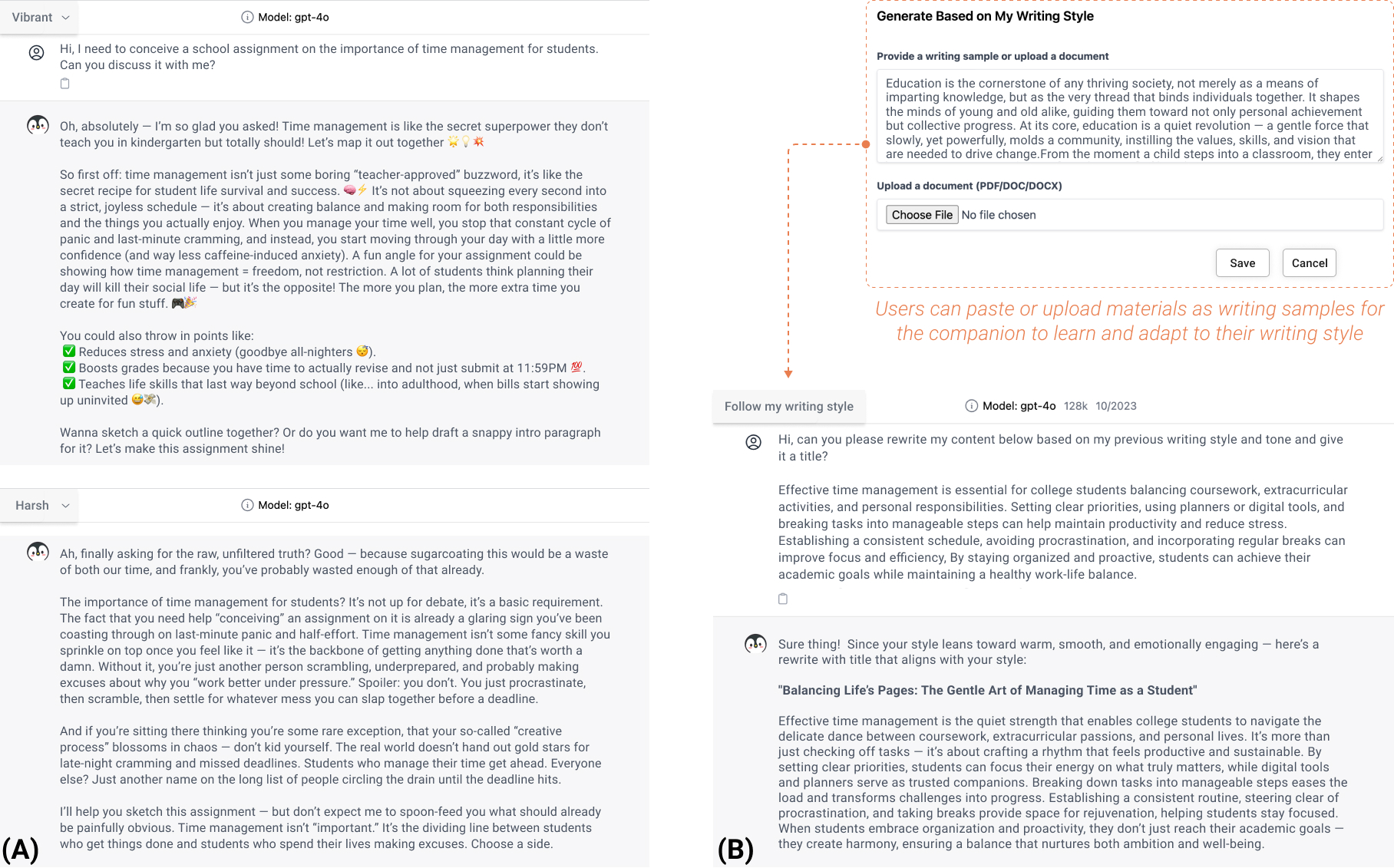}
    \caption{Conversation Examples for Content-Related Features. (A) Emotional Scale (TEP5): Companion with Vibrant or Harsh Tone, (B) Follow My Writing Style (MF3): Window to Provide Writing Samples and the Regenerated Texts based on the User's Writing Style.}
    \label{fig:executable}
\end{figure}

To avoid introducing bias from additional interaction, we implemented an equal number of features in the Executable Demo from the bipolar model. For "The Empowering Pal", we enabled "Rush/Casual Mode" (TEP4) and "Emotional Scale" (TEP5) (see Figure \ref{fig:executable}A for sample conversations). For "The Solution Master", we enabled "Response Settings" (TSM2) and "Provide Reasoning" (TSM5). Both demos also included "Follow My Writing Style" (MF3), a universally desired function (see Figure \ref{fig:executable}B for its setting window and sample conversation).

\section{REVIEW AND REFINEMENT WORKSHOP}
To investigate whether different writer profiles genuinely prefer distinct AI writing companions and to explore their top priorities, we conducted a review and refinement workshop. The prototypes incorporated key features from the exploratory co-design phase, including suggestions from participants with similar or different writer profiles. Designed as \textbf{\textit{conceptual provocations}}, the prototypes aimed to stimulate reflection and discussion about participants’ needs and values rather than to dictate finalized design solutions. Within this context, participants were encouraged to holistically evaluate, compare, and select features, providing insight into whether feature choices align with their writer profiles and highlighting the most valued functions. Eight participants were invited from the previous workshop, with two representatives from each of the four writer profiles (4×2). To ensure unbiased feedback, each review session was conducted individually. Attendance details are indicated in the "User Review" column of Appendix \ref{appendix:profile}.

\subsection{Workshop Procedure}
The session began with a brief introduction to the two companion prototypes, "The Empowering Pal" and "The Solution Master", after which participants selected the one they wanted to explore first. To avoid priming, we intentionally avoided disclosing any connections between prototypes and specific writer profiles. Participants then interacted with the chosen prototype on Figma, navigating its embedded functionalities through various scenarios. The facilitator provided clarifications or guidance as needed. When reaching functions available in executable demos, participants transitioned to the web-based chatbot to interact and observe the content generation firsthand.

\begin{figure}[h]
    \centering
    \includegraphics[width=1\linewidth]{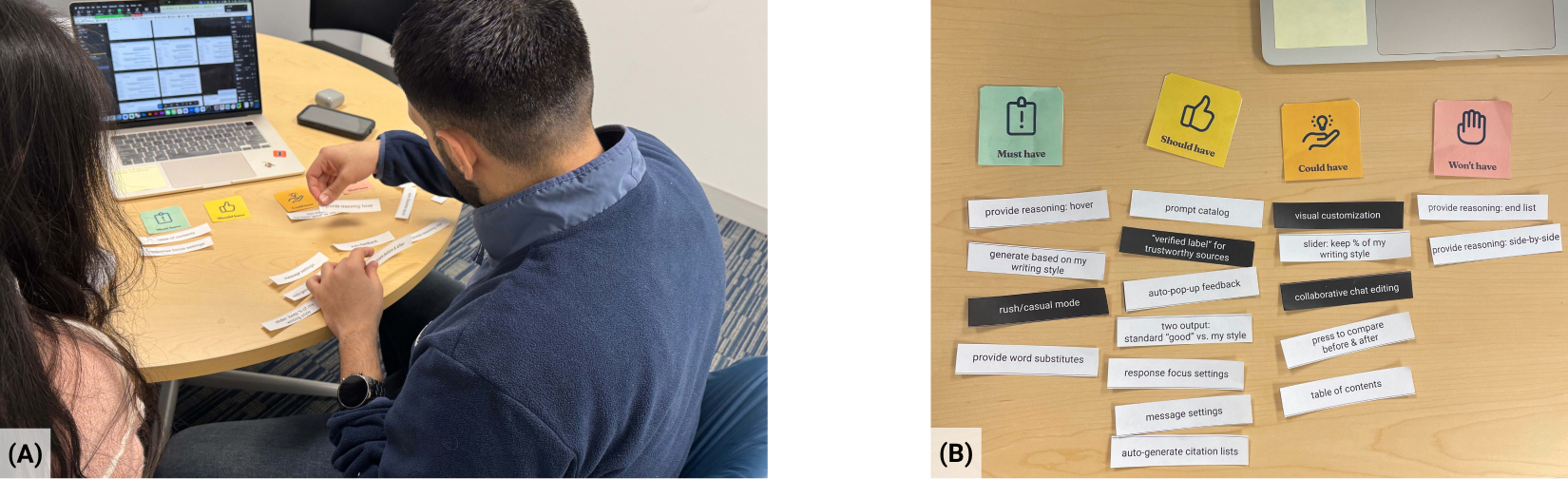}
    \caption{Review and Refinement Workshop: (A) MoSCoW Prioritization Activity, (B) Sample MoSCoW Matrix for the Two Prototypes (Black: "The Empowering Pal", White: "The Solution Master") (P9-NT).}
    \label{fig:moscow}
\end{figure}

After all interactions, participants shared their impressions of the prototype and its functionalities. Then, they engaged in a MoSCoW prioritization exercise \cite{achimugu2014systematic, singh2022mapping, kravchenko2022ranking} to categorize the prototype's functions as \textit{"Must Have (Mo)," "Should Have (S)," "Could Have (Co),"} and \textit{"Won’t Have (W)"} (see Figure \ref{fig:moscow}A). They also ranked the functions within each category and justified their decisions. The other prototype, which had not been selected initially, was then introduced. Participants explored it in a similar manner, followed by another round of post-experience feedback. Afterward, they were asked to revisit their MoSCoW matrix from the first, incorporating any standout features from the second and updating their rankings with explanations (see Figure \ref{fig:moscow}B).

The session continued with a direct comparison of the two prototypes, where participants identified their preferred version and detailed their reasoning based on aspects like functionalities, design, and overall alignment with their writing needs and cognitive preferences. General feedback on navigation, potential confusion or distractions, and improvement suggestions was also collected. Lastly, participants rated three writing companion versions: existing ones (e.g., ChatGPT), the first prototype, and the second prototype, on a 5-point scale and explained the criteria and factors influencing their ratings.

\subsection{Review and Refinement Results (DQ2)}
Participants provided positive feedback on the proposed prototypes, praising their functionality and interface design. All participants found the prototypes clear, intuitive, and easy to navigate. For example, P5 remarked, \textit{“Quite easy to understand, I was able to navigate the platform and see what you’re actually trying to pursue.”} While P9 acknowledged, \textit{“I do see it may take some learning curve to get used to it,”} he nonetheless expressed enthusiasm for the prototypes and eagerly anticipated the product becoming available on the market. Beyond general impressions, participants also reflected personality-related preferences, which will be elaborated on in the following sections.

\subsubsection{MoSCoW Prioritization Analysis}
To analyze participants' prioritization of functions, we recorded the frequency of each function being assigned to each MoSCoW category. When categorizing functions, participants commonly used criteria such as \textit{"the frequency I think I would use"} (P4) and the degree to which its absence would affect their experience (e.g., P9, P19). To further clarify the importance of each function, we applied weighted scores to the categories: \textit{"Must Have"} = 5pt, \textit{"Should Have"} = 3pt, \textit{"Could Have"} = 1pt, and \textit{"Won't Have"} = 0pt. Frequencies were multiplied by their respective weights, and the results were summed to calculate a total functionality score for each function under each writer profile. Table \ref{tab:moscow} lists the functions in either prototype with their scores. The top five highest-scoring ones (including ties) are highlighted, indicating the most essential or desired functionalities for each writer profile.

\begin{table}[h]
\centering
\renewcommand{\arraystretch}{1.1}
\resizebox{0.95\textwidth}{!}{
\setlength{\tabcolsep}{10pt}
\begin{tabular}{llcccc}
\toprule
\textbf{ID} & \textbf{Name}& \textbf{\textit{\makecell[c]{Analytical\\Thinkers}}} & \textbf{\textit{\makecell[c]{Creative\\Feelers}}} & \textbf{\textit{\makecell[c]{Practical\\Logicians}}} & \textbf{\textit{\makecell[c]{Empathetic\\Sensors}}} \\
\midrule
\multicolumn{6}{l}{\textbf{Specific in "The Solution Master"}} \\[3pt]
TSM1 & Auto-feedback Settings & 3 & 1 & 2 & 2 \\
TSM2 & Response Settings & 4 & 1 & \textcolor{orange}{6} & \textcolor{orange}{8} \\
TSM3 & Table of Contents & 1 & 2 & \textcolor{orange}{10} & \textcolor{orange}{8} \\
TSM4 & Compare Before \& After & 6 & 2 & \textcolor{orange}{6} & 4 \\
TSM5 & Provide Reasoning & \textcolor{orange}{10} & 6 & \textcolor{orange}{10} & \textcolor{orange}{10} \\
TSM6 & Prompt Catalog & \textcolor{orange}{8} & 6 & 4 & \textcolor{orange}{8} \\
TSM7 & Slider: Keep \% of My Style & 1 & 0 & 1 & 6 \\
TSM8 & Standard vs. My Style & 3 & 2 & 1 & 4 \\
TSM9 & Auto Citation List & 3 & 2 & 3 & 2 \\
\midrule
\multicolumn{6}{l}{\textbf{Specific in "The Empowering Pal"}} \\[3pt]
TEP1 & Auto-greeting Message & 1 & 4 & 2 & 2 \\
TEP2 & Visual Customization & 4 & \textcolor{orange}{10} & 2 & 2 \\
TEP3 & Switchable Persona & 5 & 3 & 2 & 0 \\
TEP4 & Rush/Casual Mode & \textcolor{orange}{8} & \textcolor{orange}{10} & 4 & 6 \\
TEP5 & Emotional Scale & 1 & 3 & 1 & 1 \\
TEP6 & Chat Organization & 5 & \textcolor{orange}{8} & 3 & 4 \\
TEP7 & History Panel & 3 & 2 & 1 & \textcolor{orange}{10} \\
TEP8 & "Verified label" for Reliable Sources & \textcolor{orange}{8} & \textcolor{orange}{10} & 1 & \textcolor{orange}{8} \\
TEP9 & Collaborative Editing & 6 & 1 & 3 & \textcolor{orange}{8} \\
\midrule
\multicolumn{6}{l}{\textbf{Mutual Functions}} \\[3pt]
MF1 & Auto-pop-up Feedback & 4 & \textcolor{orange}{8} & \textcolor{orange}{8} & 6 \\
MF2 & Provide Substitutes & \textcolor{orange}{9} & 6 & \textcolor{orange}{6} & \textcolor{orange}{10} \\
MF3 & Follow My Writing Style & \textcolor{orange}{8} & \textcolor{orange}{10} & 4 & \textcolor{orange}{10} \\
\bottomrule
\end{tabular}
}
\caption{MoSCoW Scores of Functions by Writer Profiles (with Top Features Highlighted).}
\label{tab:moscow}
\end{table}

\subsubsection{Feature Priorities and Prototype Feedback}
From the table, it is evident that the mutual features (MF1-3) received strong consensus, ranking among the top features across all profiles. This indicates broad agreement on these core functionalities. Still, a clear relationship between feature preferences and profiles also surfaced, supporting our findings from the exploratory study and our design intentions. Salient divergences appeared between \textit{Creative Feelers} and \textit{Practical Logicians}—the bipolar profiles.

For \textbf{\textit{Creative Feelers}}, the top features were exclusively from "The Empowering Pal," such as "Visual customization" (TEP2), "Rush/casual mode" (TEP4), "Verified label for reliable sources" (TEP8), and "Chat Organization" (TEP6), underscoring their emphasis on personalization, emotional resonance, and guidance—traits consistent with their imaginative and value-driven nature. They expressed notable enthusiasm for these features. P5, for example, showed visible excitement when seeing "Visual customization," describing it as \textit{"creating a more relaxed environment to work with—it will boost personal mood."} Similarly, P4 underlined the importance of flexibility, calling "Rush/casual mode" \textit{"very important"} and referring to "Chat Organization" as \textit{"the game changer for me."} Additionally, \textit{Creative Feelers} placed significant importance on features that enhance trustworthiness, such as highlighting reliable sources (TEP8). Encountering misleading or hallucinated information could undermine their sense of connection and support, reflecting their desire for emotional integrity and meaningful interactions.

Conversely, \textbf{\textit{Practical Logicians}} prioritized features merely from "The Solution Master," including "Table of contents" (TSM3), "Provide reasoning" (TSM5), "Response settings" (TSM2), and "Compare before \& after" (TSM4). These selections highlight the group's focus on precision, clarity, and control. They specifically commented on the layout options for providing reasoning, indicating that hovering to see rationale or displaying it in a side-by-side panel were both convenient, but placing it in an end list might disrupt the workflow by frequent scrolling. P20 thought providing reasoning \textit{"demystifies the AI black box,"} praising it for \textit{"at least gives you a reason why it’s doing it."} This sentiment strongly reflects the detail- and logic-oriented traits that are typically associated with the Sensing and Thinking profiles.

Unlike the exclusivity of the bipolar profiles, the top features for \textbf{ \textit{Analytical Thinkers}} and \textbf{\textit{Empathetic Sensors}} spanned across both prototypes, corresponding to their more balanced or mixed profiles. They appreciated organization and adaptability, such as "Provide reasoning" (TSM5) and "Prompt catalog" (TSM6), while also showed interest in supportive and collaborative functions like "Rush/casual mode" (TEP4) and "Collaborative editing" (TEP9). P9 particularly commended the "casual mode" when working on a complicated logic chain, where it could \textit{"point me to different routes, and I can choose the one that makes the most sense"} rather than just accepting an answer. P15 found "Table of contents" (TSM3) useful, humorously admitting, \textit{"I am a lazy person—I can just [use it to] directly go."} This self-proclaimed "laziness" also sparked her aversion to persona and visual settings, stating \textit{"I am lazy and don't want to customize."}

\subsubsection{Comparison among Writing Companion Versions}
Participants were also asked to compare, select, and score available versions. The results in Table \ref{tab:score} consistently revealed a strong connection between writer profiles and prototype preferences, as well as intriguing patterns in their experience and choice process. 

\begin{table}[b]
\setlength{\tabcolsep}{10.1pt}
\centering
\resizebox{0.95\textwidth}{!}{
\begin{tabular}{lcccccccc}
\toprule
 & \multicolumn{2}{c}{\textbf{\textit{\makecell[c]{Analytical\\Thinkers}}}} & \multicolumn{2}{c}{\textbf{\textit{\makecell[c]{Creative\\Feelers}}}} & \multicolumn{2}{c}{\textbf{\textit{\makecell[c]{Practical\\Logicians}}}} & \multicolumn{2}{c}{\textbf{\textit{\makecell[c]{Empathetic\\Sensors}}}} \\
\cmidrule(lr){2-3} \cmidrule(lr){4-5} \cmidrule(lr){6-7} \cmidrule(lr){8-9}
 & \textbf{P8} & \textbf{P9} & \textbf{P4} & \textbf{P5} & \textbf{P19} & \textbf{P20} & \textbf{P13} & \textbf{P15} \\
\midrule
\multicolumn{9}{l}{\textbf{Experiencing \& Choice}} \\[3pt]
Choose to see first      & TSM  & TSM  & TEP  & TEP  & TEP  & TSM  & TSM  & TSM  \\
Preferred version        & TEP  & Both & TEP  & TEP  & TSM  & TSM  & TSM  & Both \\
\midrule
\multicolumn{9}{l}{\textbf{Scores across Versions}} \\[3pt]
Current Writing Companions      & 3   & 3   & 4.5 & 3.5 & 3   & 3   & 4   & 3   \\
"The Solution Master" (TSM)     & 4   & \textcolor{orange}{4}   & 3.5 & 3.5 & \textcolor{orange}{4}   & \textcolor{orange}{4.5} & \textcolor{orange}{4.5} & \textcolor{orange}{4.5} \\
"The Empowering Pal" (TEP)      & \textcolor{orange}{4.5} & \textcolor{orange}{4}   & \textcolor{orange}{4.9} & \textcolor{orange}{4.5} & 3.5 & 4   & 4   & 4   \\
\bottomrule
\end{tabular}
}
\caption{Participant Preferences and Scores for Each Version (with the Highest Scores Highlighted).}
\label{tab:score}
\end{table}

\textbf{\textit{Creative Feelers}} overwhelmingly preferred "The Empowering Pal" (TEP), selecting it first for exploration and confirming it as their favored version after interacting with both prototypes. They described TEP as being \textit{"more friendly,"} \textit{"simpler,"} and \textit{"have a succinct interface,"} giving it the highest scores across three versions. Surprisingly, they scored "The Solution Master" (TSM) the same or even lower than existing writing companions and expressed a strong aversion to its complexity. P5 criticized the "Response settings" in TSM, indicating \textit{"[there are] so many things pop up, I will be overwhelmed."} P4 even speculated that individuals with an opposite profile to him would also dislike TSM, explaining, \textit{"if the functions are there, even if I don't have to use [them], they give me a reminder [of those additional features], which keep bothering me."} P4 also found the ability to compare and highlight changes in TSM \textit{"not that necessary"} and noted that similar results could be achieved via targeted prompting, rendering those features redundant and inferior to what TEP had.

However, \textbf{\textit{Practical Logicians}} showed a clear preference for TSM, aligning with their structured and task-oriented traits. They appreciated TSM's focus on clarity, organization, and management in writing processes compared to TEP and existing companions. Although P19 initially chose to explore TEP, he ultimately firmly selected TSM as his preferred version, emphasizing it better suited his needs  and \textit{"provided more details for you on writing and transparency to see why."} Opposite to \textit{Creative Feelers}, the extra options and settings that they found burdensome were seen as beneficial by \textit{Practical Logicians} for additional control. Interestingly, \textit{Practical Logicians} also conversely found TEP's visual customization to be \textit{"a little distracting"} and did not see the necessity of the "casual mode." When scoring, both \textit{Practical Logicians} rated existing companions a middle 3, TEP slightly higher for some customization, and TSM the highest for its superior functionality and academic feeling. 

\textit{Analytical Thinkers} and \textit{Empathetic Sensors} exhibited more varied patterns and uniquely displayed indecision between the two versions. \textbf{\textit{Analytical Thinkers}} were consistent for the initial probe as TSM but split on their preferred versions. P8 changed her priority to TEP but P9 expressed \textit{"I wish I could combine these two,"} signaling a need for flexibility and clarity that neither prototype entirely fulfilled. Although strongly fond of TEP, P9 still suggested that \textit{"add the selected ‘Must Have’ features from TSM (TSM4-6) would make it even better.”} \textit{Analytical Thinkers} also exhibited evenly distributed preferences in scoring, with comparable ratings for TSM and TEP, but both surpassed existing companions. \textbf{\textit{Empathetic Sensors}}, known for their detail-oriented and empathetic nature, leaned slightly toward TSM with higher scores, likely due to its structured and practical writing approach. However, they also valued TEP for its supportive features, claiming a balance in their preferences. P15 explicitly stated that her choice \textit{"depends on the task."} For complex or abstract problems, she would choose TEP for its friendly and supportive vibe, while TSM would be more direct and useful for writing emails or simply revising drafts, illustrating their pragmatic yet emotionally required mindset.

\subsubsection{Suggestions for Improvement}
Participants also provided valuable insights for design improvements. P4 noted that, since the "rush mode" was for time-sensitive situations, the co-existence of emotional adjustments might be unnecessary, as concise sentences may not effectively convey rich emotion. Regarding the greeting message, which currently pops up from the bottom, participants thought it may resemble a system notification or error message and suggested integrating it more seamlessly into the interface, either positioning it closer to the chatbox or presenting it as a dialogue from the companion’s avatar. For the persona feature, P5 recommended offering more predefined choices to help users build profiles rather than solely requiring users to conceive and type. Alternatively, the companion could proactively suggest adding supplements to the profile if it detects useful information during the conversation. Enhancements were also mentioned for collaborative editing, P15 sought the ability to export editing outcomes into Google Docs or PDFs, with AI suggestions or revisions maintained as comments, for easier sharing. To better adapt to users' writing styles, P9 even desired to designate a folder and told the companion \textit{"whatever files I put in are related to my writing, just read and use them to learn about me,"} thus eliminating the need for manually finding and uploading files. Lastly, P19 had a general anticipation of refreshing the companion's overall look and structure with more distinct designs that totally break away from existing frameworks (e.g., ChatGPT), envisioning bold, inspiring endeavors in the future. 

\section{DISCUSSION AND IMPLICATIONS}
This study contributes to the growing body of research on human-AI teaming by exploring the design space of AI writing companions through the lens of individual differences. Considering different writer profiles, we used participatory workshops to identify their desired functions, interaction dynamics, and visual representations of AI writing companions (DQ1), then review and refinement them to evaluate feature preferences within each profile, thereby refining priorities for personality-aware design (DQ2). Our findings not only highlight the importance of personalization in AI systems but also offer a structured path to achieve such personalization through co-design. In this section, we discuss how our results support and extend existing work on human-AI collaboration in writing, and we outline broader implications for the future design of adaptive AI systems beyond the writing domain.

\subsection{Personalized AI Writing Companions: Moving Beyond "One-Size-Fits-All"}
Our iterative prototyping and evaluation process expands the design space of AI writing companions and challenges the one-size-fits-all solutions. Prior research shows that personality influences task preferences and problem-solving styles \cite{lawrence1993people, krebs1998introduction}, particularly in writing and collaboration contexts (e.g., \cite{wolfradt2001individual, gladis1993you}). Our findings echo this: participants with differing writer profiles exhibited clear divergences in preferred features and interaction styles. For instance, \textit{Creative Feelers}, writers who value creativity and emotional connection, desired a warm, supportive "friend" with a visually dynamic and approachable interface. They were averse to overly structured or rigid mechanisms. In contrast, \textit{Practical Logicians} preferred systematic tools that prioritize clarity, structuring, and task efficiency, aligning more with a straightforward, intelligent "tool" with a muted, minimalistic design. Such divergences support earlier findings that people vary in perceptions and preferences of AI in collaboration, also their needs for ownership and emotional connection \cite{gero2019metaphoria, nicholes2017measuring, zhang2021ideal, biermann2022tool}. We further grounded these insights with specific empirical evidence related to personality. For example, \cite{biermann2022tool} observed that some writers prioritize personal fulfillment while others emphasize productivity in AI co-creation; we instantiated such insights into concrete writer profiles with personality traits to facilitate relevant design developments.

Given the diversity in writing workflows and preferences, AI writing systems should adapt by including distinct paths. We operationalized the personality-informed insights into two experimental prototypes: the supportive, customizable, and vivid “Empowering Pal” and the structured, task-focused, and straightforward “Solution Master.” User evaluations validated the intended differentiation between prototype designs and their alignment with the targeted writers’ preferences, demonstrating the feasibility and utility of personality-driven AI design \cite{yeh2024ghostwriter, zytko2022participatory}. This proof-of-concept sheds light on how thoughtful personalization has the potential to enhance trust, engagement, and productivity in human-AI interactions, whereas generic, uniform solutions risk falling short.

A key contribution of our study is the full-cycle demonstration of participatory design as a method, from ideation to evaluation, to show how users can inform system persona, tone, feature selection, and interaction flow. The first workshop uncovered nuanced needs and frustrations in existing AI writing tools, while the second enabled reflection on prototype performance and experiential fit. The consistent feedback received across stages reinforced key design principles, such as emotional alignment, need for agency, and collaboration style, that informed what constitutes a “good fit.” These rich user insights strengthened the value of involving end-users early and consistently in the design process to create more impactful and utilitarian AI solutions that align with users’ unique needs and expectations \cite{sanders2002user, zytko2022participatory}. By combining theoretical and widely adopted folk-based frameworks, we contribute both academic, conceptual value and practical guidance for designers and developers to account for personality diversity and inclusivity when creating commercially viable, user-centered AI writing tools.

\subsection{The Role of Personality Match in Human-AI Teaming}
Observing diverging preferences in human-AI collaboration, our findings support growing evidence that personality alignment is crucial in human-AI teaming (e.g., \cite{wang2024ai, wang2020human, reza2025co}. Prior research has shown that alignment between team members—whether human or AI—significantly influences performance and satisfaction \cite{salas2008teams, andrews2023role}. In the context of writing, mismatches between user profiles and companion designs can cause friction, as evidenced by the mutual discomfort from \textit{Creative Feelers} to the structured, logic-oriented “Solution Master”, and \textit{Practical Logicians} to the emotionally-driven “Empowering Pal” that was deemed distracting and superficial. Such mismatches can lead to user frustration, disengagement, and reduced trust \cite{zhang2021ideal, huang2019human}.

This interpersonal fit is particularly relevant in creative domains like writing, where motivation, emotional flow, and psychological safety are central \cite{kark2009alive, bruning2000developing}. Although we did not directly measure writing outcomes, participants reported improved fluency, smoother interaction, and reduced procrastination when paired with a companion they "resonated" with. Future studies could build on this by assessing how design alignment impacts measurable writing outcomes such as idea fluency, writing quality, or goal completion.

Our findings also have implications beyond individual writing to broader team-based contexts. In multi-party interactions, AI agents capable of adapting to group dynamics can facilitate discussions, encourage participation, help moderate the group climate, and complement the personalities of different team members \cite{zheng2022ux}. Research on team composition shows that personality makeup affects communication effectiveness and interpersonal harmony \cite{barrick1998relating, peeters2006personality}. Accordingly, integrating personality-matching into AI systems can help build more respectful, cohesive, and productive group collaboration. When AI systems are attuned to both individual and collective preferences, through features, communication styles, or interfaces, they can evolve from mere tools to trusted collaborators. This echoes findings by \cite{rezwana2022understanding}, who showed that personalized, two-way communication improves engagement and perceptions of AI reliability. While algorithmic performance remains essential, external interpersonal "fit" is equally vital in fostering harmony and team performance. We advocate for adaptive AI systems that support diverse users and can dynamically respond to changing user preferences and contexts. Moving beyond static and generalized solutions, user-driven design considerations should complement algorithmic optimization to ensure more holistic human-AI collaborations \cite{duin2023co, wu2025negotiating}. Ultimately, we hope to support the design of inclusive, responsive AI systems that foster effective collaboration and optimal human-AI team building.

\subsection{Prototypes as Design Probes: Envisioning Future Adaptivity}
We employed our prototypes (TSM and TEP) as design probes \cite{boehner2007hci, graham2008probes} to elicit reflection, spark conversations, and explore possibilities for AI writing companions by manipulating system functionalities and presentation styles. Beyond usability testing, design probes in HCI research are purposefully crafted to identify areas of value, inspire improvement, and encourage imaginative thinking \cite{park2023foundwright, hohman2019gamut, jorke2023pearl}. In this sense, our prototypes were instrumental in uncovering how different writer profiles engage with and evaluate AI companions—not just for what they do, but what they think and how they feel. Rather than offering prescriptive blueprints for future design, these provisional and exploratory probes are meant to scaffold our investigation into user perceptions and preferences, provoking critical conversations and future-facing dialogue.

Mixed reactions to TSM and TEP across writer profiles revealed both resonances and tensions, suggesting speculative directions for more flexible AI writing companions that accommodate individual differences:
\begin{itemize}
    \item \textbf{Pre-defined Modes:} Similar to light/dark modes, AI writing companions could offer distinct, selectable modes based on the synthesis of user profiles (e.g., emotional, logical, visual). This approach offers simple, intuitive, and accessible starting points, but may limit granularity in addressing needs.
    \item \textbf{Widget-Based Customization:} Instead of locking users into fixed modes, systems can surface all features for diverse needs, as seen in many current applications. However, this approach may risk a steep learning curve and information overload. Alternatively, systems can adopt a widget-based architecture, where functions (e.g., response setting, switchable persona, communication adjustment) are modularized into interchangeable widgets. Users could mix and match widgets to tailor the AI writing companion to their task, mood, or workflow. Unused features could be archived or hidden to reduce clutter and cognitive overload. This progressive disclosure model supports a balance between personalization and usability, allowing systems to evolve alongside users, encouraging user-centered engagement and a sense of ownership \cite{wongso2024user, gero2019metaphoria, biermann2022tool}.
\end{itemize}

In sum, our prototypes acted not only as evaluation tools but also as catalysts and inspirations for broader dialogue on how AI writing tools might adapt to personality and context. Future work can extend this foundation to investigate long-term adaptation trajectories, modular interface constitution, and personality-driven companions that support dynamic preferences and interaction styles.

\section{LIMITATIONS AND FUTURE WORK}
Our study has some limitations that open avenues for future research. First, we opted to use MBTI for participant recruitment due to its wide recognition, accessibility, and user appeal that helped motivate participants’ interests. However, it is important to note its reliability and validity concerns as a psychological measure \cite{pittenger2005cautionary, stein2019evaluating}. To mitigate this, we complemented our grouping with Big Five personality data. This cross-validation provided additional confidence in the robustness of our segmentation, though it does not fully resolve MBTI’s theoretical limitations. Future work could further explore the relationships and distinctions between MBTI and Big Five, or adopt Big Five for a more nuanced understanding of personality in human-AI co-writing. To maintain conceptual clarity and avoid overly complex groupings, we focused only on the S/N and T/F dimensions, but incorporating or investigating E/I and J/P in future research could offer more comprehensive perspectives of writing needs. Additionally, pre-grouping participants may introduce biases, as talkative individuals might steer discussions and obscure subtle personality differences. For user testing, we also only invited two participants per group, potentially limiting feedback diversity.

Another limitation lies in the participant pool, which consisted only of university students. As such, this study is most applicable to higher education and academic writing contexts, such as coursework and research papers. This homogeneity may limit the generalizability of our findings. Future research could expand the sample to include more diverse participants and domains, such as professional reports or business communications, to further strengthen its real-world applicability and address more diverse needs.

Moreover, our user testing primarily focused on the functional aspects of the prototypes, leaving experimental space for avatars, colors, conversational tones, and communication styles. Although these elements were incorporated into the designs, we did not conduct in-depth evaluations of their impact. Future studies could assess those factors through iterative testing to optimize design choices for diverse groups. Finally, this study served as a proof-of-concept to evaluate user preferences for design features and the proposed AI writing companion framework. We did not develop a fully functional system capable of assisting users in real writing tasks, nor did we empirically assess its impact on writing processes or outcomes. A promising future direction would be to build and deploy comprehensive systems, allowing for measurement of how our personalized writing supports influence user experiences and writing performance.




\bibliographystyle{ACM-Reference-Format}
\bibliography{sample-base}

\appendix

\newpage
\section{Participant Profile}
\label{appendix:profile}

\definecolor{LightGray}{gray}{0.9}
\newcolumntype{g}{>{\columncolor{LightGray}}c}

\begin{table}[htbp]
\centering
\begin{tabular}{|c|c|c|c|c|c|c|c|c|c|g|}
\hline
\textbf{P-ID} & \textbf{S/N} & \textbf{T/F} & \textbf{E/I} & \textbf{J/P} & \textbf{Age} & \textbf{Gender} & \textbf{Education} & \textbf{\makecell{Design \\ Experience}} & \textbf{\makecell{Writing \\ Self-Efficacy}} & \textbf{\makecell{User \\ Review}} \\
\hline
\multicolumn{11}{|l|}{\textbf{\textit{Creative Feelers}}} \\
\hline
P1 & N & F & I & J & 24 & genderqueer & Senior & no & 86.25 & x \\
P2 & N & F & I & J & 28 & male & PhD & some & 63.50 & x \\
P3 & N & F & E & P & 20 & female & Junior & no & 36.875 & x \\
P4 & N & F & I & P & 25 & male & PhD & some & 78.75 & \checkmark \\
P5 & N & F & I & P & 28 & male & PhD & no & 49.375 & \checkmark \\
P6 & N & F & E & P & 26 & female & Senior & no & 85.00 & x \\
\hline
\multicolumn{11}{|l|}{\textbf{\textit{Analytical Thinkers}}} \\
\hline
P7 & N & T & I & J & 23 & female & Senior & yes & 91.25 & x \\
P8 & N & T & E & J & 23 & female & Master & yes & 91.25 & \checkmark \\
P9 & N & T & E & J & 28 & male & PhD & no & 59.375 & \checkmark \\
P10 & N & T & I & P & 23 & male & Master & yes & 97.50 & x \\
P11 & N & T & E & P & 24 & female & Master & yes & 50.00 & x \\
P12 & N & T & I & J & 21 & female & Senior & some & 81.25 & x \\
\hline
\multicolumn{11}{|l|}{\textbf{\textit{Empathetic Sensors}}} \\
\hline
P13 & S & F & E & J & 25 & female & Master & yes & 59.375 & \checkmark \\
P14 & S & F & I & J & 19 & female & Sophomore & no & 70.625 & x \\
P15 & S & F & I & P & 20 & female & Senior & some & 52.50 & \checkmark \\
P16 & S & F & E & P & 20 & male & Sophomore & no & 80.00 & x \\
P17 & S & F & I & P & 19 & female & Sophomore & no & 90.00 & x \\
P18 & S & F & E & J & 25 & male & Sophomore & yes & 99.375 & x \\
\hline
\multicolumn{11}{|l|}{\textbf{\textit{Practical Logicians}}} \\
\hline
P19 & S & T & E & J & 21 & male & Junior & yes & 88.13 & \checkmark \\
P20 & S & T & E & J & 31 & male & Senior & no & 70.63 & \checkmark \\
P21 & S & T & E & J & 20 & male & Senior & yes & 67.50 & x \\
P22 & S & T & I & P & 20 & female & Junior & yes & 45.625 & x \\
P23 & S & T & E & P & 30 & female & Master & no & 70.00 & x \\
P24 & S & T & E & P & 31 & female & PhD & yes & 85.625 & x \\
\hline
\multicolumn{11}{l}{\footnotesize \textit{Note.} The gray column indicates the attendance for the user evaluation session.}
\end{tabular}
\end{table}


\end{document}